\documentclass[acmlarge, screen]{acmart}

\copyrightyear{2021}
\acmYear{2021}
\setcopyright{acmlicensed}
\acmConference[COMPASS '21]{ACM SIGCAS Conference on Computing and Sustainable Societies (COMPASS)}{June 28-July 2, 2021}{Virtual Event, Australia}
\acmBooktitle{ACM SIGCAS Conference on Computing and Sustainable Societies (COMPASS) (COMPASS '21), June 28-July 2, 2021, Virtual Event, Australia}
\acmPrice{15.00}
\acmDOI{10.1145/3460112.3471951}
\acmISBN{978-1-4503-8453-7/21/06}

\begin{document}

%%
%% The "title" command has an optional parameter,
%% allowing the author to define a "short title" to be used in page headers.
%\title{\sysnames: Transformer-based Multi-scale Indoor Human Mobility Modeling using Passive Sensing}
\newcommand{\sysnames}{WiFiMod}
\newcommand{\sysname}{WiFiMod }
\newcommand{\shorttitle}{WiFiMod: Indoor Human Mobility Modeling using Passive Sensing}

\title[WiFiMod: Indoor Human Mobility Modeling using Passive Sensing]{\sysnames: Transformer-based Indoor Human Mobility Modeling using Passive Sensing}

%%
%% The "author" command and its associated commands are used to define
%% the authors and their affiliations.
%% Of note is the shared affiliation of the first two authors, and the
%% "authornote" and "authornotemark" commands
%% used to denote shared contribution to the research.

%\author{Anonymous Authors}

\author{Amee Trivedi}
\affiliation{%
  \institution{University of Massachusetts Amherst}
  \country{USA}
}
\email{amee@cs.umass.edu}

\author{Kate Silverstein}
\affiliation{%
  \institution{Oracle Labs}
  \country{USA}
}
\email{kate.silverstein@oracle.com}

\author{Emma Strubell}
\affiliation{%
  \institution{Carnegie Mellon University}
  \country{USA}
}
\email{strubell@cmu.edu}

\author{Mohit Iyyer}
\affiliation{%
  \institution{University of Massachusetts Amherst}
  \country{USA}
}
\email{miyyer@cs.umass.edu}

\author{Prashant Shenoy}
\affiliation{%
  \institution{University of Massachusetts Amherst}
  \country{USA}
}
\email{shenoy@cs.umass.edu}

%%
%% By default, the full list of authors will be used in the page
%% headers. Often, this list is too long, and will overlap
%% other information printed in the page headers. This command allows
%% the author to define a more concise list
%% of authors' names for this purpose.
\renewcommand{\shortauthors}{Trivedi, et al.}

%%
%% The abstract is a short summary of the work to be presented in the
%% article.
\begin{abstract}
Modeling human mobility has a wide range of applications from urban planning to simulations of disease spread. It is well known that humans spend ~80\% of their time indoors but modeling indoor human mobility %has not received much attention as compared to outdoor mobility modeling due 
is challenging due to three main reasons: (i) the absence of easily acquirable, reliable, low-cost indoor mobility datasets, (ii) high prediction space in modeling the frequent indoor mobility, and (iii) multi-scalar periodicity and correlations in mobility. %Additionally, most of the work on modeling models mobility at the same granularity of the dataset, which xxx. 
To deal with all these challenges, we propose \sysnames%\footnote{We have publicly released the code of \sysname under BSD for use by other parties at http:\textbackslash\textbackslash<url anonymized for blinded submission>}
, a Transformer-based, data-driven approach that models indoor human mobility at multiple spatial scales using WiFi system logs. % and models indoor mobility at multiple spatial granularities using transformers.
\sysname  takes as input enterprise WiFi system logs %generated across an enterprise network 
to extract human mobility trajectories from smartphone digital traces. Next, for each extracted trajectory, we identify the mobility features at multiple spatial scales, \emph{macro} and \emph{micro}, to design a multi-modal embedding Transformer that predicts user mobility for several hours to an entire day across multiple spatial granularities. Multi-modal embedding captures the mobility periodicity and correlations across various scales while Transformers capture long term mobility dependencies boosting model prediction performance. This approach significantly reduces the prediction space by first predicting \emph{macro mobility}, then modeling indoor scale mobility, \emph{micro mobility}, conditioned on the estimated macro mobility distribution, thereby using the topological constraint of the macro-scale. %Additionally, transformer captures long term mobility dependencies boosting model prediction performance. 
Experimental results show that \sysname achieves a prediction accuracy of at least 10\% points higher than the current state-of-art models. %We further present 2 applications of our model - a COVID-19 policy making and resource allocation application on occupancy prediction application to predict the occupancy of zones inside a building to alert the authorities for possible hot pockets that violate the COVID19 occupancy policy and a personalized model to cater to individual human mobility. 
Additionally, we present 3 real-world applications of \sysname - (i) predict high density hot pockets and space utilization for policy making decisions for COVID19 or ILI, (ii) generate a realistic simulation of indoor mobility data to simulate spread of diseases, (iii) design personal assistants.
\end{abstract}

%%
%% The code below is generated by the tool at http://dl.acm.org/ccs.cfm.
%% Please copy and paste the code instead of the example below.
%%



%\ccsdesc[500]{Human-centered computing~Smartphones}

%%
%% Keywords. The author(s) should pick words that accurately describe
%% the work being presented. Separate the keywords with commas.

%%
%% This command processes the author and affiliation and title
%% information and builds the first part of the formatted document.
\maketitle

\section{Introduction}
\label{sec:introduction}

Understanding human mobility is fundamental to location based services, urban transportation, and smart cities among many other applications and  paramount to improve sustainability.  %through various datasets over the years.
Lately, with the advances in networking technologies and ubiquity of mobile phones, a large amount of mobility data is generated and collected in the form of GPS logs, cellular data, social media check-ins, and vehicular data giving rise to data-driven human mobility modeling \cite{hang2018exploring, veloso2011urban, hasan2013understanding,jurdak2015understanding}. This prior work seeks to capture human mobility at urban scales \cite{isaacman2012human} using transportation, social media, and phone data. While taxi or public transit data \cite{Ganti:2013:IHM:2493432.2493466, veloso2011urban} allow urban-scale mobility of users to be captured from a vehicular or transportation standpoint, social media check-in data \cite{jurdak2015understanding} enables users' mobility to be tracked at various points of interest \cite{ hasan2013understanding}. GPS and cellular data from phones have also been used to capture urban mobility patterns, with GPS capturing fine outdoor mobility and cellular data capturing coarse mobility \cite{huang2010activity}. However, all these modeling efforts focus on outdoor or \emph{macro scale} human mobility across various Point of Interest (POI), locations, or city regions. 

Studies have shown that humans spend over 80\% of their lives indoors \cite{arif2016impact} resulting in indoor or \emph{micro mobility}. Recent research has recognized that indoor mobility of users inside buildings,  where many users spend a significant portion of the day, is very different from outdoor mobility exhibited when walking in a city or traveling in vehicles \cite{Zheng:2018:BAM:3276774.3276780,zhou2017mining}. As we model mobility at a finer spatial scale, mobility becomes more frequent and the prediction space expands. We argue that the motivation for indoor mobility, as well as the region of movement, is time-dependent and micro mobility shows high correlations to the macro mobility features of context, location type, and location name. %resulting in a hierarchical approach. 
Moreover, indoor mobility displays a complex sequential periodicity correlated to the macro, outdoor or coarse grained, features of mobility. Due to the above stated reasons, we cannot directly use outdoor mobility models that capture mobility at large grid or POI levels at a single spatial scale for indoor mobility modeling. %More importantly most of the prior work needs indoor localization, which is time consuming and requires costly human labor, or installation of sensors for acquiring indoor mobility dataset. we found that there is an absence of easily acquirable, cheap, and reliable data source to use for indoor mobility modeling. 

In this work, we present \sysnames, a transformer-based multi-scale indoor mobility model that uses existing WiFi infrastructure to passively sense human mobility. In pursuit of this model, we have three specific goals. First, we argue that human mobility is inherently hierarchical, where macro mobility, user type, and time of the day determine the micro mobility. %For eg., the probability of a student visiting an educational building for a lecture is high during work hours at 11 am than at 11 pm. Also, we argue that a user's future micro mobility conformance %%or deviation to their past micro mobility 
%is influenced by the type of building visited by the user. % such as an educational building or dining hall further determines the conformance or non-conformance of user micro mobility inside the building. %likelihood of conformance behavior of the user to visit the same location as in the past. 
%For eg. users display conformance to their past micro mobility when they visit an educational building type to attend a lecture by visiting the same indoor room and display non-conformance to past micro mobility when they visit a dining hall for lunch or dinner.% that a student will visit the has a high ctime of the day and the context of mobility (work, home, leisure) determines location type visited (cafeteria, conference rooms, library,etc) further location type determines the location or building name and then narrows down to the indoor space label such as room number or floor number in the building. 
Second, we capture the multi-modal features of macro as well as micro mobility patterns by creating a joint embedding and learn the correlations to generate sequences of context (Work or Home), building type (describes the space usage), building name (unique building identifier), and indoor location (room number, floor, or zone). The multi-modal embedding captures how individuals move between indoor spaces across and within buildings and takes into account how different space types exhibit distinct mobility patterns over time due to differences in space utilization. Third, we provide a ready-to-deploy system that uses existing ubiquitous WiFi infrastructure present at all enterprise networks and uses system log (\emph{syslog}) messages to extract indoor human mobility. %In today's era, humans carry their mobile phones everywhere with them and on a campus/enterprise network WiFi is ubiquitous making WiFi logs a very reliable data source for indoor human mobility. We use the syslogs generated across Access Points (AP) on an enterprise network that passively log all network activities from mobile devices on the network and extract indoor human mobility trajectories across buildings over time. %Most mobility models fall short on one or all of the above goals. Primarily, detecting real-world indoor mobility is hard. 
 %A growing body of research has shown that WiFi based human mobility is accurate \cite{}. WiFi syslogs provide information on the spatio-temporal components of the user mobility trajectory and the spatial components helps us derive the context, location type, outdoor and indoor location attributes.

%From a systems standpoint, we find that the current efforts in modeling human mobility cater to designing models at the same spatial granularity of the dataset and this results in a loss in learning the mobility correlations across multiple spatial granularity. %coarser macro granular mobility to finer micro mobility. 
%From the application perspective, each level of the multi-scale mobility modeling gives insights to a different application design so understanding and choosing the right spatial scale for modeling is very important from the systems design perspective as demonstrated in the case studies. For eg, indoor hot pocket identification for de-densification of areas as we reopen from COVID19 need indoor level micro mobility modeling. However, designing a campus layout to identify the placement of different building types needs modeling mobility at a building type macro level.

Our main contributions in this work can be summarized as follows:
\begin{itemize}
    \item We design an end-to-end data-driven approach to model indoor human mobility using passive WiFi sensing. WiFi logs based passive sensing approach uses already existing WiFi infrastructure in an enterprise or campus network providing a reliable indoor mobility dataset.
    \item We propose the use of multi-modal embedding to capture the macro and micro mobility features along with their correlations to improve the model prediction accuracy.
    \item We demonstrate the efficacy of our model by evaluating it against a real world dataset of 2500 users in a large campus setting and show that our model shows superior performance by at least 10\% points over other indoor mobility models.
    \item We present three case studies that demonstrate the use of our model in predicting indoor hot pockets or high human density zones, generating user mobility trajectories, and designing personal assistants. % with a real world campus dataset.
    %\item Indoor human density prediction for policy making.
    %\item location type space utilization distribution over time.
    %\item conduct extensive experiments and show
    %\item We show 2 applications COVID19 and personalized model... I think just talk about COVID19 mainly no personalized model.
\end{itemize}

\iffalse
Para 1 
\\
- human mobility modeling is important has multiple applications. justify why.
\begin{itemize}
    \item Motivate the current problem of 80\% time spend indoors so modeling indoor mobility is important and outdoor indoor mobility is quite different : more frequent. So we can't use the outdoor models.
    \item and current models do mobility modeling at the same granularity as that of the dataset.
    \item mobility is hierarchical in nature, so we need to model everything
\end{itemize}
Para 2: what we need to do and what has been done. Limitations of current work.
\\
Para 3 : Contributions of our work
\begin{itemize}
    \item produce indoor mobility model to accurately model the hot pocket areas for contagious disease spread like COVID19
    \item aggregate movement to produce movements of individuals to reproduce human densities over time to spot locations, context and activities over time at geographical scale
    \item different granularities exhibit different mobility pattterns we want to capture the hierarchical nature of mobility.
\end{itemize}
\fi

\section{Background and User Data Ethics}

\begin{figure}
\includegraphics[width=0.81\linewidth]{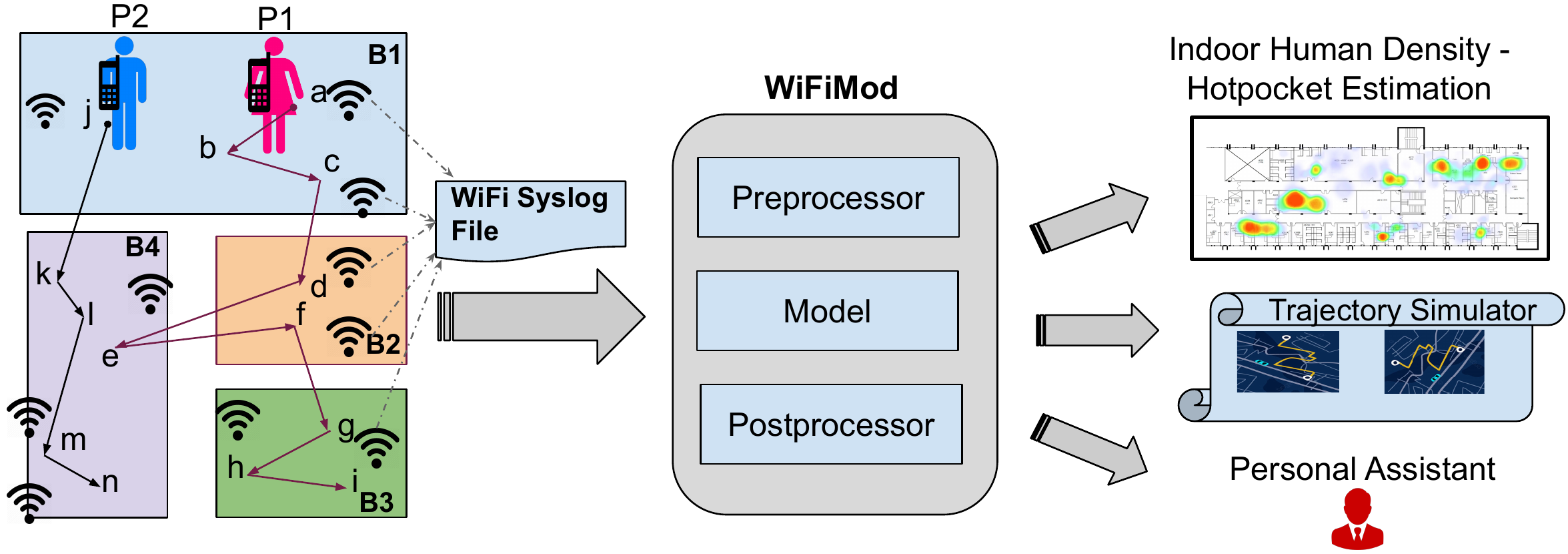}
\caption{WiFiMoD : An indoor mobility modeling approach using WiFi sensing. \label{fig:WiFiMoD_Figure}}
\end{figure}

%\begin{figure*}[t]
%\begin{center}
%    \begin{tabular}{cc}
%    \includegraphics[width=0.72\linewidth]{Figures/Hierarchical_Mobility_user_view_context_space_outdoor_indoor.pdf} &
%    \includegraphics[width=1.2in]{Figures/placeholder.png}\\
%(a) & (b) 
%\end{tabular}
%\end{center}
%%\vspace{-0.2in}
%    \caption{(a) Mobility Hierarchical View (b) PDF of locations visited across different spatial scales}
%\vspace{-0.05in}
%\label{fig:hierarchy}
%\end{figure*}

In this section, we present the background for our work on data-driven indoor mobility modeling.

\textbf{Mobility as Nomadic Behavior:}
Some mobility models, such as the classic random waypoint model, emphasize modeling the physical movement of users such as velocity, acceleration, and direction of movement \cite{camp2002survey, davies2000evaluating}. In contrast, several other models, including our work, view user mobility as inherently nomadic. Nomadic user mobility can be seen as a sequence of location visits, where users visit a location to spend some time at that location known as a {\em dwell} period then {\em transition} to another location, followed by a dwell period at the new location and so on \cite{qiao2018hybrid}.
Figure ~\ref{fig:WiFiMoD_Figure} shows two users P1 and P2 visiting multiple buildings B1 through B4 and spending time at various locations. Each dwell period at buildings B1 through B4 for P1 and P2 is followed by a transition. In this case, the emphasis is on \emph{which} locations are visited at various times of the day, across multiple buildings, building types, and context, revealing the semantic meaning of the nomadic behavior. %.what is the {\em time of visit} to a location, {\em how long} does the nomadic device or user stay at that location and {\em what path} does the nomadic device or user take to transition between locations. 
Since humans are creatures of habits and tend to follow a routine \cite{song2010modelling}, we need to capture the correlations such as repeating visits to a location, repeating sequences resulting from daily or weekly routines, long-term dependencies, and affinity to certain locations, to name a few. While transitions from one location to another also need to be modeled, the emphasis is on capturing nomadic behavior, rather than factors such as the speed of mobility, the direction of movement, mode of transport, etc. Since, our primary focus is on modeling indoor mobility, modeling nomadic behavior is more appropriate since users are often stationary inside the building - in their office, in meetings, etc.

\textbf{Modeling Trajectories:}
Mobility models come in many different flavors depending on what aspects of mobility the model is attempting to capture. A common type of mobility modeling to capture nomadic behavior is next location prediction \cite{Do:2012:CCM:2370216.2370242, liu2016predicting, Lin:2012:PIM:2370216.2370274, gambs2012next, gidofalvi2012and, mathew2012predicting} where the model attempts to predict the next location that will be visited by the user. Next location prediction can be used in mobile systems for location-aware services, caching, etc. In contrast, our modeling approach focuses on modeling and predicting the entire trajectory of the user (and devices) over the next few hours to an entire day. Modeling and predicting trajectory over many hours or entire day can be viewed as a generalized and more complex problem than next location prediction, since, doing so involves predicting a long sequence of future location and not just the next one. A trajectory is essentially a temporally ordered sequence of locations visited, duration of stay at each location, with transitions between two successive locations where the transit is the path used to move from the previous location to the next one.
Figure ~\ref{fig:WiFiMoD_Figure} shows the trajectory of users P1 and P2 as a sequence of locations each visited for a specific time duration at a certain time of the day. Modeling the entire trajectory provides a holistic view of how users and devices move throughout a day.% and has many applications to system designing including optimizing HVAC in building based on user visits and occupancy \cite{trivedi2017ischedule}, recommendation systems based on recurring location visits, etc.

%\textbf{Human Mobility}
%Para 1
%\begin{itemize}
%    \item Define mobility, context, location type, location name, indoor location name
%    \item define trajectory
%    \item define the periodicity, temporal and spatial attributes of mobility
%\end{itemize}
%\textbf{Mobility Modeling}
%Need for indoor mobility modeling
%\\

\textbf{Modeling Different Spatial Scales:} A key design consideration in indoor mobility modeling is the spatial scale for capturing the nomadic movement of users and devices. Generally, models are designed to capture mobility or nomadicity at a single spatial scale and this spatial scale is often the same as that in the underlying dataset used to derive the models. For example, cellular data sets have been used to model mobility at the spatial scale of cell towers. In this work, we argue that indoor mobility models should be capable of modeling nomadic movement at different spatial scales and the choice of which spatial scale to choose should depend on what higher-level problems need to be solved using the model. While some prior work has focused on context-aware modeling they do not take into consideration the multiple spatial scales of mobility \cite{Liao:2018:PAL:3304222.3304245, Do:2012:CCM:2370216.2370242}.

In the case of indoor mobility within and across buildings, at least two spatial scales are desirable from a modeling perspective. For models that are derived using WiFi traces, the finest spatial scale for nomadic movement is that of an Access Point (AP), which roughly translates to mobility at the scale of a room or a group of rooms in the span of a single AP. This spatial scale reveals {\em micro-scale} nomadic movement inside each building. It is also useful to consider coarser spatial scales such as considerably larger spatial regions (e.g. an entire floor) as a single location and consider nomadic movement across such coarser spatial regions. Another useful spatial scale is to consider an entire building as a single coarse-grained location to model {\em macro-scale} nomadic movement. In this case, a trajectory comprises visit to buildings, time spend inside a building, visit time of buildings, and transitions between buildings; at this scale, we are only concerned with which building (e.g. in a university campus) users visit and not how they move inside that building. %Figure ~\ref{fig:macro_micro}(b) shows the macro and micro mobility timeline of user P1 and P2. We see that the macro mobility timeline shows the mobility at building level as sequence of buildings visited whereas, micro mobility timeline shows mobility inside each of the buildings visited in the macro mobility trajectory along with the time of visit and duration of visit at each of the indoor location inside each building.

Different spatial scale models lend themselves to solving different types of problems. For example, a macro-scale model is useful for designing location-aware recommendations when a user visits a building, while a micro-scale mobility model is useful for indoor resource scheduling and hot pocket identification. %adding AP placement optimization in wireless network design or optimizing HVAC schedules based on room-level occupancy patterns. %We argue that multiple spatial scales are best modeled using a hierarchical approach.
As noted earlier, we employ a hierarchical approach for modeling mobility at multiple scales. Doing so not only enables our models to predict both macro- as well as micro-scale
mobility patterns, it is also more efficient---it reduces the prediction space by first predicting mobility patterns at the macro scale and then modeling micro scale patterns conditioned on the estimated macro scale patterns.

%\textbf{Impact of Spatial Scale on Mobility}
%\begin{itemize}
%    \item mobility is more frequent as granularity increases... 1 plot on this (PDF of locations visited, unique/non-unique)
%    \item argument on why mobility is hierarchical, how to show the correlation... how to motivate this? Should I show some co-relation/scope of locations? What plot to show here...
%    \item main figure example elaboration
%\end{itemize}

\textbf{WiFi Logs Based Passive Sensing:}
Today, WiFi is ubiquitous at university campus, enterprise, and urban locations. When users move across the campus with their mobile devices, the devices get associated and disassociated with access points (AP) along the user's mobility route. These device associations and disassociations get logged as events into the system log, syslog, of each AP. We use the AP syslog file to passively observe the user devices as they move across the network and derive user mobility by using the smartphone as an alias for user mobility since users carry their mobile phones with them everywhere. The key benefits of using WiFi syslog for passive sensing are (i) we do not need any new installation or deployment of any devices as, in most places, the WiFi syslogs are collected by the Information Technology (IT) department to analyze network performance or network attacks; (ii) no data collection on the user device needs to be done and no user intervention is needed to collect the data, and (iii) WiFi is present indoors and thus WiFi logs provide a viable method to learn indoor mobility.

\begin{figure}[t]
\includegraphics[width=0.81\linewidth]{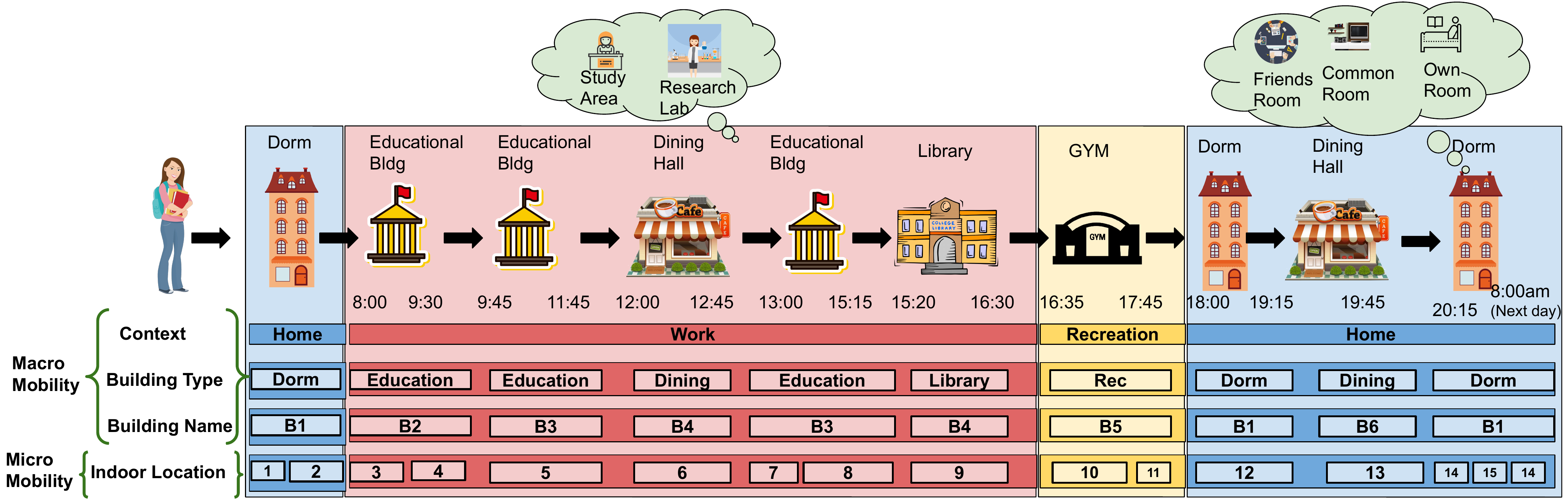}
%\vspace{-0.2in}
\caption{Mobility Hierarchical View}
\vspace{-0.05in}
\label{fig:hierarchy}
\end{figure}

\textbf{Multi-Scale Mobility}
While it has been shown that user mobility displays recurring patterns at a scale, we argue that human mobility is inherently hierarchical, where hierarchy is represented by spatial granularity scale as it becomes fine grained micro mobility from a coarse grained macro mobility representing context, building type, and building name. %Here, we present some empirical analysis for the same. 
As shown in Figure \ref{fig:hierarchy}, a user who visits several locations to accomplish their daily tasks seems extremely mobile at the scale of indoor location, visiting 14 locations throughout the day. As we change the spatial granularity to a coarser grain, we find that the mobility becomes infrequent at the building scale, where the user visits 10 buildings. Finally, at the context level---which defines the overall span of activities the user performs in the part of the day---the user shows mobility across only 4 contexts. Thus, showing that human mobility becomes more frequent as the spatial scale becomes fine grained. Also, each indoor location space shows high affinity to the context and building type displaying dependencies and correlations between macro and micro scale mobility features. 

\begin{figure*}
\begin{center}
    \begin{tabular}{ccc}
    \includegraphics[width=1.5in]{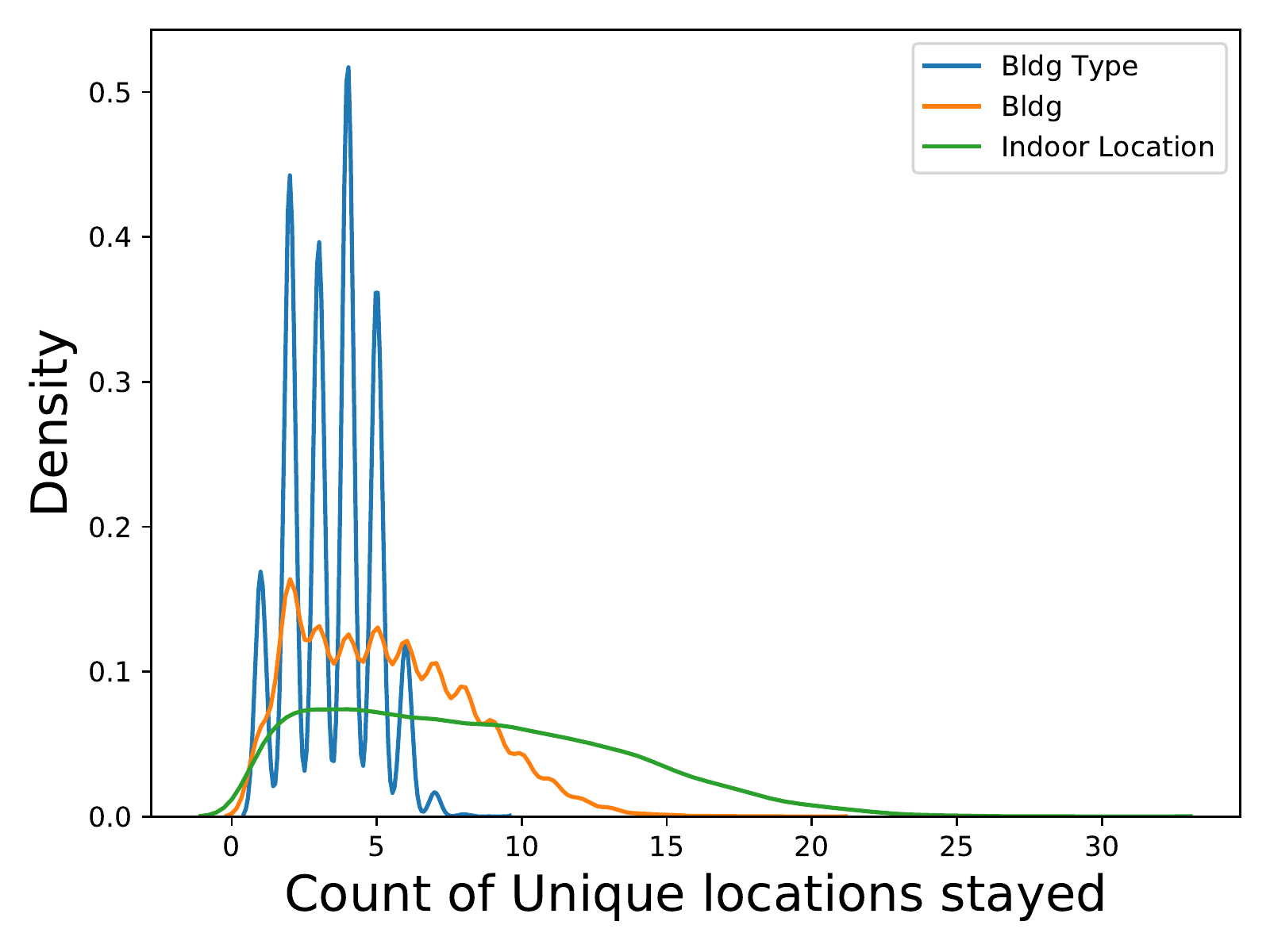} &
    \includegraphics[width=1.5in]{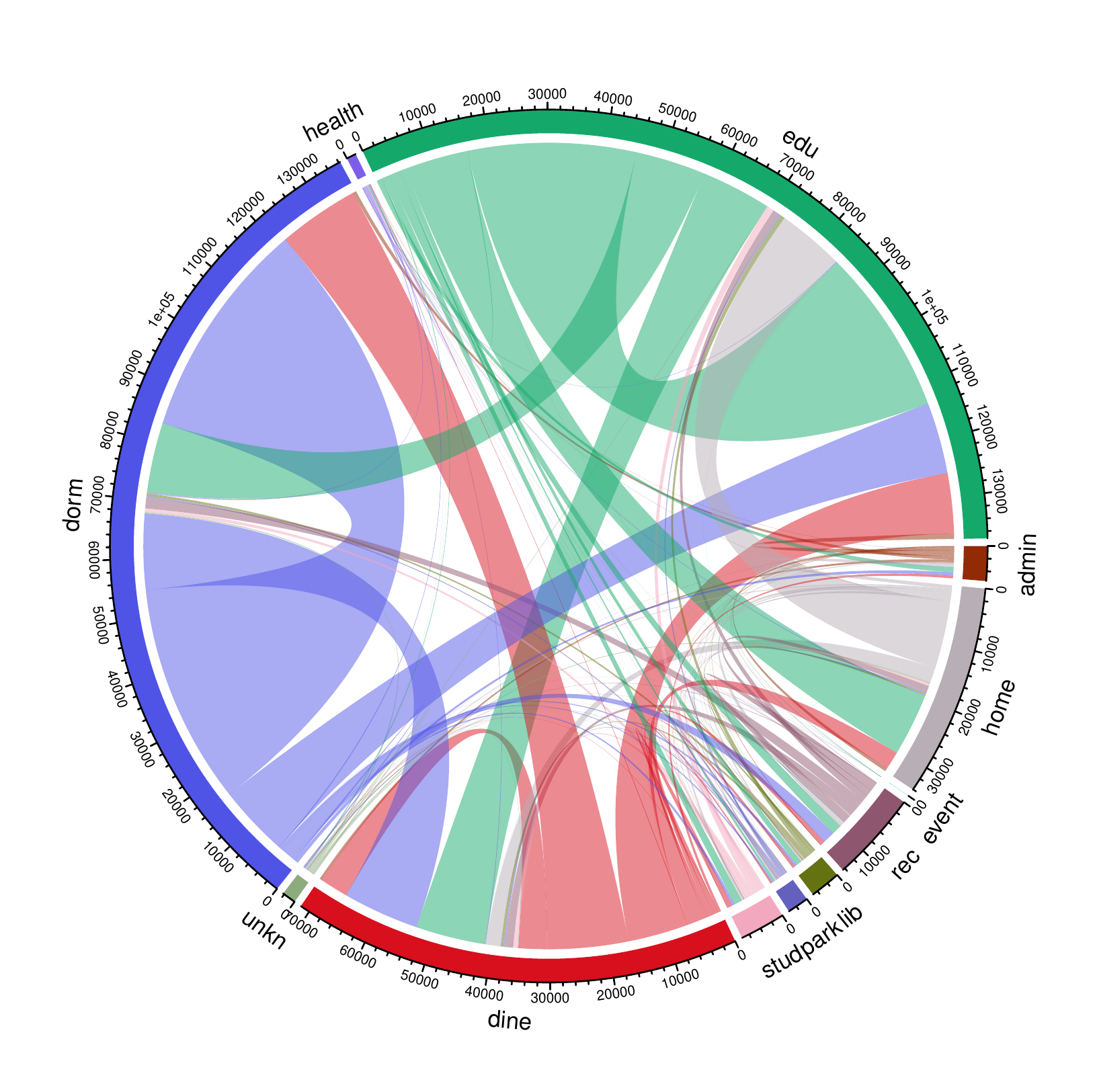}  &
    \includegraphics[width=1.5in]{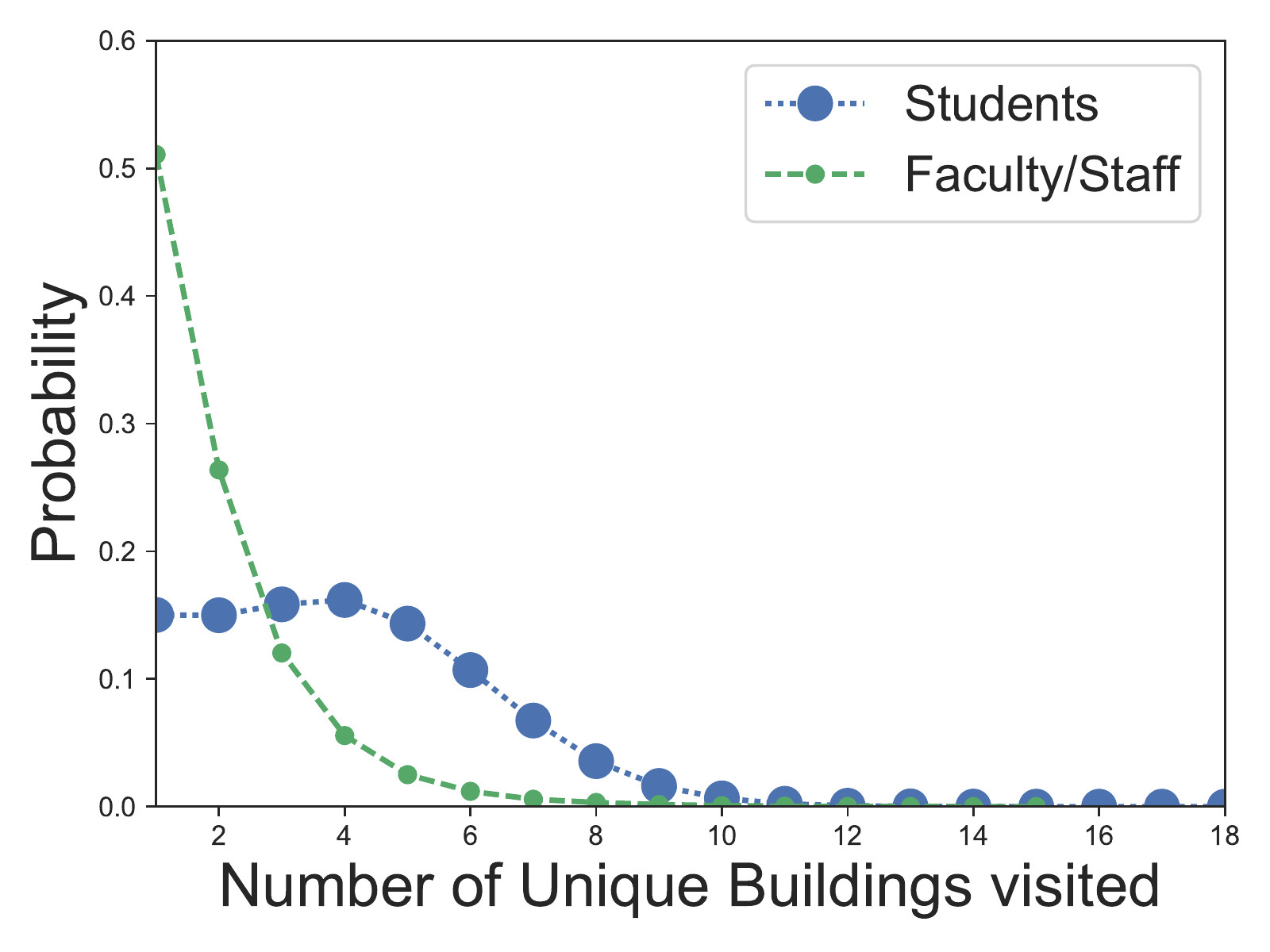} \\
(a) &(b)&(c)
\end{tabular}
\end{center}
%\vspace{-0.2in}
    \caption{Features influencing indoor mobility prediction (a) Spatial Scale (b) Building Type (c) User Type
    }
\vspace{-0.05in}
\label{fig:indoor_factors}
\end{figure*}

\textbf{Features Impacting Indoor Mobility Prediction}
We conducted empirical analysis on a large campus WiFi syslog dataset described in \S \ref{sec:evaluation} and found that four main factors impact micro mobility:
\begin{itemize}
    \item \textbf{Spatial Scale:} Figure \ref{fig:indoor_factors}(a) is a density plot of count of dwell locations of users across an entire day at each spatial scale. Dwell location is defined as a location where users spend at least 10 minutes. \emph{Context} describes the situational factors such as work or home. \emph{Building type} indicates the building usage activity: for example, a food court is used for dining, while a building with classrooms is used for education. \textit{Building name} is the location name visited, and the \textit{indoor location} is the location inside the building visited as shown in Figure \ref{fig:hierarchy}. We see that the average number of visits are 4, 5, and 11 at building type, building name, and indoor location level respectively. Giving us the insight that \emph{as the spatial scale becomes more fine-grained, from context to indoor mobility, the user mobility becomes more frequent.}
    \item \textbf{Building Type:} Figure \ref{fig:indoor_factors}(b) is the chord diagram showing user movements within and across different building types. We see that an educational building, as well as dorms, see more dwell locations within the buildings where other building types such as admin, and dining see relatively less within building dwell locations. The main reason is that students move from one classroom to another within and across educational buildings during work hours resulting in a high number of dwell locations in education buildings. %for this is the that students spend more amount of time in educational and dorm buildings. 
    This indicates that the \emph{space type that governs the primary activity inside the building plays is an important feature in indoor human mobility}. 
    \item \textbf{User Type:} our campus dataset has two types of users, students, and faculty, as identified by the role field in the authentication events of syslog messages. Figure \ref{fig:indoor_factors}(c) shows the distribution of the unique number of buildings visited by users (students and faculty/staff), here multiple visits to a building count as a single unique location. We see that on an average a faculty/staff visits 1.2 unique buildings per day while students visit an average of ~3 unique buildings per day. Thus, illustrating that user type influences the observed user mobility.
    \item \textbf{Past Behavior:} We find that the future mobility of a user is highly dependent on past behavior. Users who display high conformance behavior in the past continue to do so in the future. This observation is inline with the findings in prior work \cite{jayarajah2018predicting}.
\end{itemize}

%\begin{figure}
%\includegraphics[scale=0.3]{Figures/placeholder.png}
%\caption{PDF of locations visited at different spatial scale. %\label{fig:pdf_scale}}
%\end{figure}

\subsection{Ethical Considerations}
\label{sec:ethics}
Our study has been approved by our Institutional Review Board (IRB) and is conducted under a Data Usage Agreement (DUA) with the campus network IT group that restricts and safeguards all the WiFi data collected. To avoid any privacy data leakage all the MAC ids and usernames in the syslogs are anonymized using a strong hashing algorithm. The hashing is performed before syslog data is stored on disk under the guidance of the IT manager who is the only person aware of the hash key of the algorithm. Any data analysis that results in the de-anonymization of the users is strictly prohibited under the IRB and signed DUA. All users using the campus WiFi network need to provide consent to the campus IT department for syslog data events from their devices to be stored for a system diagnosis or analysis of attacks on the enterprise network. Additionally, all researchers sign a form of consent to adhere to the signed IRB and DUA and undergo mandatory ethics training.
\section{Problem and Approach}
\label{sec:system}

\subsection{Problem Statement}
We focus on the problem of modeling indoor mobility trajectories of users over the timescale of several hours to a day. We assume that historical indoor mobility data for each user is available for purposes of modeling. A trajectory of a user over a duration such as a day is defined to be a sequence of tuples (c,s,b,l), where each tuple comprises of context (c), space type (s), building location name (b) and indoor location name (l). Our model seeks to predict the trajectory of each user while learning the correlation between the c,s,b, and l at multiple spatial granularities. Further, we model trajectories inside a single building as well as those that span a collection of nearby buildings.

\begin{figure}
    \includegraphics[width=0.9\linewidth]{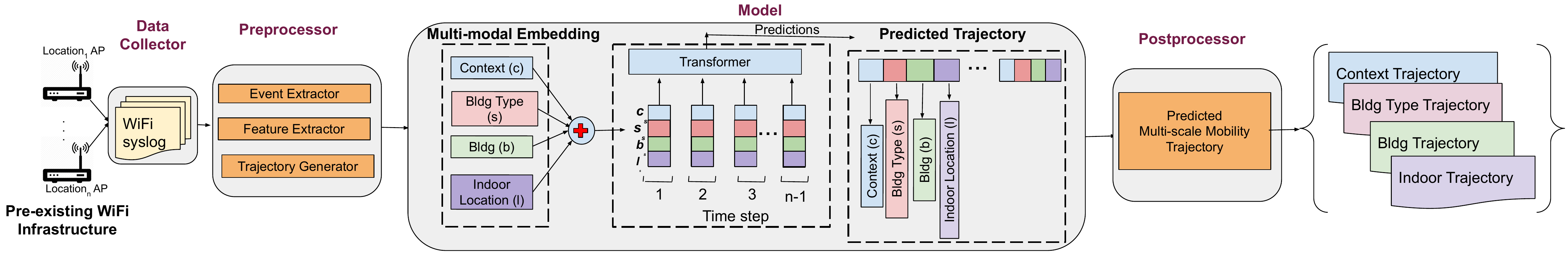}
%\vspace{-0.2in}
    \caption{System Architecture Diagram}
\vspace{-0.05in}
\label{fig:sys_arc}
\end{figure}
\subsection{System Overview}

Figure \ref{fig:sys_arc} shows the architectural overview of \sysnames. \sysname is a pipeline of 3 main modules: data collector, data preprocessor, and model. The main objective of the data collector is to collect the WiFi syslog files across all the APs in the enterprise network. Most IT departments already have the networking logging turned on; if disabled then the IT admins would need to turn “on” the network logging to enable data collection. The output of this module is an aggregated syslog file from all APs across the campus. The aggregated syslog file is fed to the data preprocessor, which extracts the events and fields needed to generate user trajectories from the raw syslog files. This module is vendor-specific, depending on the vendor of the deployed AP. Currently, \sysname supports HP-Aruba syslog files. Once the user trajectories are extracted they are fed into the model, which extracts the macro and micro mobility features, creates a multi-modal feature embedding, and feeds it to the Transformer model. The output of the model are predictions, which can be used to generate reports or are aggregated to predict space usage and occupancy for various applications.

\subsection{WiFi-Based Modeling Approach}
A large, campus-like infrastructure  comprises of various building types such as dormitory, educational, dining, student union, research labs, health center, recreational center and administrative. Campus users move across multiple buildings everyday to accomplish their tasks and use resources scattered across campus. A campus or enterprise WiFi network provides seamless WiFi coverage inside buildings and between buildings through Access Points (AP) installed across the geographical area of the institution. As users move within this geographical area, their devices connect and hop  APs. Each AP maintains an internal log that consists of a list of all events observed by the AP. %To provide ubiquitous WiFi to thousands of users throughout the campus a large number of WiFi access points are deployed across all locations. 

When a user connects their device, it associates with a nearby AP. Each AP has a fixed location identified by the room, floor and building of installation. As a user moves across multiple locations on the campus, the device gets associated and disassociated with multiple APs on the user’s path. The association and disassociation events, along with timestamp, Device MAC, AP ID, and event type get logged in the internal syslog file maintained by each AP. Extracting all the association, disassociation or drift events from syslog files of all APs on the campus and indexing them by timestamp gives us a sequence of APs visited and duration of visit by each user device. Since all AP locations are known in terms of building, level and room of installation, it further helps us derive user device trajectory information at multiple spatial scales. % To derive the context, we use a simple heuristic where all activities between 8:30am-5pm are marked as “work" context and rest as “home” context because of the university timings.

The enterprise WiFi network on campus is operated with RADIUS authentication that mandates all users to authenticate before connecting to the network. Since today’s users carry a plethora of mobile devices we extract these authorization messages from syslog files to create a user-to-device map and use this to identify the mobile devices (typically the smartphone) of each user and use its trajectory as an alias of the user trajectory. 

Now, to train a data-driven model, we collect syslogs and extract trajectories for each user for a few weeks and create a historic trajectories dataset for training the model. From the extracted trajectories, we derive macro and micro mobility features based on the building type and heuristic rule for context defined above. This serves as the input to the model, which is a global model trained on all user trajectories. We use the multi-level spatial features of each trajectory to create a multi-modal embedding and train the Transformer. The predictions of this model are then used as is for individual mobility or can be aggregated.

\subsection{System Architecture}
\subsubsection{Preprocessor:}
The syslogs collected from the APs are a deluge of data mainly used for system diagnosis or analysis of attacks on the enterprise network. A typical syslog is a collection of diverse timestamped events, where each event has a pre-specified format. The goal of the preprocessor is to extract the relevant events from the syslog file and convert the events into a trajectory. The preprocessor is a sequence of 3 main steps: event extraction, data dependency resolution, and trajectory generator. 

In the first step, we extract association, disassociation, reassociation, authorization, deauthorization, and drift event messages, hereby referred to as presence messages, from the syslog file. The event format is as shown below:
\begin{verbatim}
       <Timestamp> <hh:mm:ss> <controller_name> <event_id> 
       <message_body : MAC_ID , AP_ID, other text>
\end{verbatim}

The timestamp field gives us the time of event; $event\_id$ gives us the event type; $message\_body$ consists of device $MAC\_ID$, which identifies each device uniquely, and $AP\_ID$, which gives us the AP details namely building name, level and room number. Authorization and deauthorization messages additionally have username and role fields that help create a mapping between users and their devices, used for selecting the most mobile device from the collection of devices owned by each user, along with the role of the user on campus identified as student or faculty/staff. 

The event logging in syslog has lots of inconsistencies such as dropped events, time sequence events overlap, multiple similar events, incorrect order of events, multiple disparate event types logged for the same device at the same timestamp, to name a few. Such inconsistencies need to be resolved before the mobility trajectory of the device aka user is computed. The main objective of this step is to resolve these inconsistencies, estimate the missing entries, clean the data, and generate a timestamped sequence of rows of association and disassociation of devices with AP. 

After that, we gather all events per user device and create a timestamp indexed sequence to identify the APs visited, along with the time of visit, to generate a mobility trajectory. Then, for each generated indoor mobility trajectory, we add the corresponding context, building type, and building name to each visited indoor location. We generate the context based on a simple heuristic that campus working hours are between 8:30am and 4:30pm, so all user activities between these times are marked as "work" context and the rest are marked as "home" context. We find that students who stay on-campus display both these contexts whereas for off-campus users, we generally see only the work context except for students in research labs who work outside the work context hours and students who stay on campus to use recreation and student union facilities later or early during the day. Each building on our campus has a specific usage assigned to it (e.g. educational building have classrooms, dining has food courts, recreational building has swimming pools, squash courts, gymnasium). We use the designated space activity as the location space type. Thus, for each indoor location visited in the extracted WiFi trajectory we compute the corresponding context, space type, and building name resulting in a sequence of (c,s,b,l) tuple as the multiple spatial granularity trajectory.

%perform 2 main steps before feeding them to the transformer model. In the first step it generates the context, building type and building name from each item in the trajectory sequence for every device. We generate the context based on a simple heuristic that campus working hours are between 8:30am and 4:30pm, so all user activities between these times are marked as work context and the rest are marked as home context. We find that students who stay on-campus display both these contexts whereas for off-campus users we generally see only the work context except for students in research labs who work outside the work context hours and students who stay on campus to use recreation and student union facilities later or early during the day. Each building on our campus has a specific usage assigned to it for eg - educational building have classrooms, dining has food courts, recreational building has swimming pools, squash courts, gymnasium, etc- we use the designated space activity as the location space type. ***

%%%%%% NEW VERSION
\subsubsection{Transformers for Sequential Prediction:}
The Transformer neural network architecture \cite{vaswani2017attention}, originally introduced for the task of machine translation, follows an encoder-decoder structure. The encoder maps a sequence of inputs $\bf{x}$ consisting of the inputs $x_i$ at each position $i$ to a sequence of continuous representations $\bf{z}$. These representations are provided as input to a decoder that autoregressively generates an output sequence of labels $\bf{y}$, with the prediction at each output timestep conditioned on the entire input sequence $\bf{z}$. The length of the output sequence is not tied to the length of the input sequence. In the Transformer architecture, the encoder and the decoder share the same neural network architecture structure, except that in the decoder, the representation at position $i$ is prevented from observing representations at subsequent positions. We describe this architecture in more detail below. 

First, a sequence of input tokens is first mapped to corresponding $d_{model}$-dimensional input embeddings via an embedding lookup table. These embeddings are then fed to the encoder of the Transformer, which is comprised of $L$ layers of the same form. Each layer $j$ passes its inputs through two sub-layers, multi-head self-attention and a feed-forward layer, with residual connections (addition followed by normalization) between each:

\begin{align*}
    h_1 &= MultiHeadAttention(\bf{x}^{(j-1)}) \\
    h_2 &= LayerNorm(\bf{x}^{(j-1)} + h_1) \\
    h_3 &= FeedForward(h_2) \\
    \bf{x}^{(j)} &= LayerNorm(h_2 + h_3)
\end{align*}

%Self-attention first performs three linear projections on the representation at each position, creating a query, key, and value representation for each position. The quer
For the representation $x_i$ at a given position in the sequence, self-attention computes scores between $x_i$ and every other representation in $\bf{x}$, and uses those scores to compute a weighted average (attention) over the representations at all positions. In multi-head self-attention, this operation is performed $k$ times, so that $k$ different attention functions can be learnt, to model different dependencies between elements in the sequence. For more low-level details, see \cite{vaswani2017attention}.
%is basically a softmax distribution for each word in the input including itself. 
For each attention head, three matrices $Q, K, V$ are created by multiplying the input (a sequence of embeddings) with weight matrices $W^Q, W^K, W^V$ (of dimension $d_{model} \times d_{model}$, $d_{model} \times d_k$, and $d_{model} \times d_v$, respectively).\footnote{In practice, it is common to set $d_k = d_v = d_{model}$, as we do in this work.} Using the terminology from \cite{vaswani2017attention}, $Q$ represents "queries", $K$ represents "keys", and $V$ represents "values". The multi-head attention mechanism allows for the model to jointly attend to information from different representation subspaces along different positions. Layer normalization is applied after residual connections to improve optimization. 

\[ 
MultiHead(Q, K, V) = Concat(head_1, ..., head_h) W^o
\]

where,
\[head_i= Attention(QW_{i}^{Q},KW^{K}_{i},VW^{V}_{i})
\]

Attention is given by the formula:

\[ 
Attention(Q,K,V) = softmax \left( \frac{Q K^T}{\sqrt[2]d_k} \right) V
\]
%$Attention(Q,K,V) = softmax(QKT√dk)V$

% and parameter matrices
% \[W^Q_i \in R^{{d_{model}}\times d_k},W^K_i \in R^{d_{model}\times d_k},W^V_i \in R^{d_{model}\times d_v} and  W^O \in R^{{hd_v}\times d_{model}}
% \]

It is also worth noting that, unlike recurrent neural networks (RNNs) or convolutional neural networks (CNNs), Transformers are not inherently sensitive to sequential order, and it is therefore common to inject information about the position of elements in the input sequence via \textit{positional encodings}. These positional encodings are added to the input representations.

\subsubsection{Multi-Modal Transformer Model:} 
To learn the sequential as well as long-term dependencies from the  input trajectories we use a Transformer-based autoregressive (sometimes called "causal") language model. We use an off-the shelf Transformer implementation based on GPT-2 \cite{Radrof2019} and train it from scratch on our dataset.\footnote{We use the GPT-2 \cite{Radrof2019} implementation available in the HuggingFace Transformers \cite{wolf-etal-2020-transformers} library, version 4.4.2.} We treat the task of predicting the next set of locations visited by the user as a task of language modeling, where language modeling task is defined as the task of predicting next character or word in a document.

Our Transformer model takes as input $m = 4$ trajectories corresponding to the \textit{context, space type, building}, and \textit{indoor location} spatial modalities generated by the preprocessor. We map the events in each raw trajectory to an index in a shared vocabulary of size $V$, to obtain $m$ integer-valued sequences $T_1, ..., T_{m}$, each with length $n$. Then, we map the entries in each sequence to learned, $d$-dimensional \textit{event embeddings} $E_1, ..., E_m$, where $E_i = \langle e_1, ..., e_n \rangle$ and each $e_i \in \mathbb{R}^d$. Since the vocabulary is shared, we use a separate set of event embeddings for each modality to avoid collisions when event ids from different modalities happen to overlap. Since Transformer models inherently lack an inductive bias that would allow them to be sensitive to different sequential orderings, we also learn $d$-dimensional \textit{position embeddings} $P$ for each of the $n$ positions. We obtain a single joint embedding by summing together $Je = E_1 + ... + E_m + P$. The derived joint embedding $Je$ is passed through $m$ stacked Transformer encoder layers, each of which has $m$ attention heads. 

We train this model using a self-supervised autoregressive training objective: given the events at timesteps $1, ..., i-1$ in each modality, the model is trained to predict the events that occur at timestep $i$. In other words, the model estimates $p(c_i, s_i, b_i, l_i | c_{1:i-1}, s_{1:i-1}, b_{1:i-1}, l_{1:i-1})$. To make predictions, we pass the outputs obtained from the Transformer encoder through an additional linear layer of dimension $d \times V$, which is shared across all modalities. We hypothesize that using a shared output layer encourages the embeddings for different modalities to maintain a coherent geometry relative to one another; however, we leave in-depth analysis of different architectural choices for future work.

During training, we convert the logits obtained from the output layer to (log) probabilities via the softmax function, then compute the batchwise-mean cross entropy loss for each modality. We sum these together to obtain a final combined loss. We minimize this objective over $15$ epochs using the Adam optimizer \cite{Kingma2015} with a learning rate of $0.01$ and a mini-batch size of $100$. As mentioned above, we use $m = 4$ Transformer layers and $m = 4$ attention heads per layer. We set the embedding dimension $d$ to $64$.

\section{Experimental Evaluation}
\label{sec:evaluation}

%Datasets and Parameters
%\\
%Baseline Comparison and Metrics
%\\
%Hierarchical v/s Single Tier
%\\
%Top 1/3/5 prediction
%\\
%Space Type prediction
%\\
%15/30/60 mins prediction accuracy
%\\
%Covid19 indoor Hotpocket Prediction
%In this section we present the evaluation results of our model performance as compared to the existing state of art models. We also evaluate the prediction effectiveness and robustness of our system across all tiers at various granularities. We have released the source code, implementation details and a tool ready to be used for modeling user/device mobility at https://github.com/ **anonymized for double-blind review**

\subsection{Dataset and Parameter Setting} \label{sec:metrics}
%Describe dataset and put a table with consolidated dataset description about inter and intra building values, dwell times, device ownerships. 1 plot if needed else only table

\textbf{Dataset} For the evaluation of our model we use campus-scale device trajectory dataset extracted from WiFi logs of a large university campus(name removed for double-blind review). Table \ref{tab:dataset_stats} provides dataset details. Our campus comprises of 156 buildings spread over 1463 acres and has seamless wireless connectivity through 5104 HP Aruba access points (AP). These APs are managed by seven wireless controllers and they receive syslog messages of all events seen by the APs. AP logs contain many types of events, of which six events types are relevant to our study: association, disassociation, reassociation, authentication, de-authentication and drift events. Since the campus operates an enterprise WiFi network with RADIUS authentication, all user devices must authenticate themselves before they connect to the network. Doing so generates authentication and deauthentication log messages, which allows the network to associate each device with a particular user. Once authenticated, the device can then associate with a nearby access point, which generates an association message in the event logs. If the device moves out of range or wakes up from sleep, it may generate deassociation, reassociation or drift message. Each event in the log consists of a timestamp, device MAC ID and Access Point ID. In addition, authentication and deauthentication events also include the user ID. For privacy reasons, all device MAC ID and user ID are anonymized using a SHA-1 hash function as noted in section \S\ref{sec:ethics}. Since the location of all access points are known (in terms of the building and floor where they are deployed), each of these event types represents a {\em "presence"} message. The sequence of presence messages generated by a device over the course of the day reveals all the AP (and building-specific) locations visited by that device and the time spent at each location. Further, since each device must first authenticate to the network with the user ID for RADIUS authentication, the owner of each device is known, which in turn reveals the collection of devices owned by each user. As noted earlier, this data has been collected, and anonymized, under an IRB protocol approved by our Institutional Review Board. We remove stationary network devices identified by association with a single AP for the entire dataset duration and select only the most mobile device (smartphone) as an alias for the user mobility. Identification of devices owned by each user and selection of only the most mobile device as alias of user mobility helps avoid double counting users as well. For evaluation of \sysname we use event log for the 2 months of Fall'19 is over 150GB in size and contains 6.4 billion events.

\begin{table}
    \begin{center}
     \begin{tabular}{lc} \toprule
     Item Description & Value \\ \midrule
     Number of Users &  2000 \\%30084 \\
     Number of Building Types & 13 \\
     Number of Buildings & 156 \\
     Number of APs &  5104 \\
     %Number of Devices & \\%61326 \\
     %User Type & Students, Faculty and Staff \\
     %Avg. Devices per user &  2.03\\
     Avg. Buildings visited & 4 \\ 
     Avg. Indoor locations stayed & 8.97/building \\
     %Avg. Inter-bldg dwell time & 109 mins \\ 
     Time Span  & Fall'19: Sep-Nov '19\\ \bottomrule
     \end{tabular}
     \caption{Dataset Description}\label{tab:dataset_stats}
 \end{center}
\vspace{-0.3in}
 \end{table}

\textbf{Parameter Setting: } To evaluate the robustness of our proposed model we use a train-dev-test split of 80-10-10 where we use the first 80\% data of each user as training data, next 10\% as dev and rest 10\% as testing data. For the selection of model hyper-parameters, we use a grid search over the parameter space and select the optimal parameter settings using the dev dataset. Parameter optimization is performed using mini-batch Adam optimizer and with a batch size of 100. %All the following experiments have been performed with a 2 layer seq2seq RNN with 128 hidden units in each layer and a dropout rate of 0.2. 

\subsection{Baseline Comparison}
To evaluate the effectiveness of our model we compare our proposed model with the following:

% 1) ARIMA 2) n-gram [bigram and trigram with and without smoothing] \cite{} 3) Hidden Markov Model (HMM) \cite{} 4) Periodic Mixture Model (PMM) \cite{} 5) ST-RNN \cite{} 6) Gmove \cite{}. 

\textbf{N-gram: } An n-gram model is one of the most important tools in speech, language and text processing. An n-gram model is used to estimate the conditional probability of visiting a location given the sequence of previously visited locations. We include evaluations against first and second order Markov chains as the baseline. A bi-gram model uses past location to estimate the probabilities (using MLE), whereas tri-gram approach conditions on past 2 locations.

%\textbf{ARIMA: } ARIMA \cite{} is a classic statistical method for learning and predicting values with time series data. It is an integrated generalization of the auto-regressive moving average (ARMA). We use ARIMA to model the location sequences of the user mobility by building a mathematical model with historic data to represent the regular patterns of a time series dataset and forecast the future values of the constructed time series. %The values of p,d and q for the model are p=5,d=2 and q= 5.

\textbf{HMM: }  In a Hidden Markov Model (HMM) we regard all visited locations as state and build a transition matrix based on the sequence of locations visited. We train one HMM for all users and each hidden state generates locations over a Gaussian distribution. 

%\textbf{PMM} A Periodic Mobility Model (PMM) assumes that mobility follows a spatiotemporal mixture model and takes periodicity into consideration when predicting the next location.

%\textbf{ST-RNN} An ST-RNN model is a simple RNN model without any attention mechanism. The ST-RNN models local temporal and spatial contexts in each layer with time-specific transition matrices for different time intervals and distance-specific transition matrices for different geographical distances. This is a current state-of-the-art technique.

\textbf{LSTM} Long Short Term Memory (LSTM) has shown superior performance for sequential data and encoding long term dependencies, so we use LSTM as one of our baselines.

\textbf{Simple Transformer} is an adaption of our model, which does not perform multi-modal embedding. For indoor modeling we train the simple transformer with the historic indoor trajectories. It is a basic autoregressive language model implemented with Transformers.
%\textbf{GMove} GMove is the state-of-the-art method based on HMM. It constructs several HMMs and assigns users to each HMM by a soft label proportional to the probability of drawing the trajectory from the HMM.
%\vspace{-0.31in} 
\begin{table}
    \begin{center}
    \begin{tabular}{cccc} \toprule
        Model Name & 15 mins & 30 mins & 60 mins \\ \midrule %#Location Accuracy (Inter) & Location Accuracy (Intra) &
        bi-gram  & 5.32\% & 6.24\% & 7.94\%\\ 
        tri-gram  & 9.51\% & 12 .3\% & 19.24\% \\
        four-gram  & 14.45\% & 17.5\% & 24.8\% \\
        %bi-gram with Kneser-Ney smoothing & & & & \\ 
        %tri-gram with Kneser-Ney smoothing  & & & & \\
        %ARIMA   & &  \\
         HMM  & 15.16\% & 20.21\% & 25.36\% \\ 
            %PMM  & & & & \\
       %ST-RNN  & & & & \\
        LSTM & 57.48\% & 62.3\% & 71.96\% \\ %\textbf{71.96\%} & 37.49\% & \\
           %GMove  & & & & \\ \bottomrule
        %Simple  Transformer &  & & \\
        Simple Transformer & 64.28\% & 69.93\% & 75.81\% \\
        Our Model  & \textbf{68.39\%} & \textbf{79.4\%} &  \textbf{83.2\%}   \\ \bottomrule
    \end{tabular}
     \caption{\sysname indoor mobility prediction comparison with Baseline Models.}\label{tab:baseline}
 \end{center}
\vspace{-0.35in} 
\end{table}

\iffalse
\begin{figure*}[t]
    \begin{tabular}{p{2.1in}p{2.1in}p{2.1in}}
%\includegraphics[width=2.2in]{./figures/Ground_Truth.png} &
%\includegraphics[width=2.2in]{./figures/Model_accuracy.pdf}&
\includegraphics[width=2.1in]{Figures/tier1_duration_ctaj_nohue_450.pdf} &
\includegraphics[width=2.1in]{Figures/tier2_duration_nohue_1.pdf} &
\includegraphics[width=2.1in]{Figures/tier3_duration_ctaj_nohue_450.pdf} \\
    (a) Tier-1 [Macro scale] Strong Pearson correlation coefficient: 0.76 & (b) Tier-2 [Indoor/Micro scale] Strong Pearson correlation coefficient: 0.69  & (c)Tier-3 [Multi-device] Strong Pearson correlation coefficient: 0.78
\end{tabular}
\vspace{-0.1in}
\caption{Joint plot of predicted duration and observed duration across the 3 tiers of the hierarchical model.}
\vspace{-0.05in}
\label{fig:duration}
\end{figure*}
\fi

\textbf{Results:} Table \ref{tab:baseline} shows the comparison results between our proposed model and the baseline models. For the evaluation, we predict the entire indoor trajectory generated by each model for each user at a temporal granularity of 15 mins, 30 mins, and 60 mins and check the predictions against the ground-truth locations to compute the model accuracy. We evaluate \sysname against other baselines %on next location predictions at indoor location prediction 
and find that Transformer-based \sysname outperforms both the LSTM model and HMM. Transformers have a higher-order transition modelling capacity than a HMM. In addition, the multi-head self-attention mechanism allows it to capture long-term dependencies more effectively than an LSTM. 

In general, the deep neural network (DNN) based models show superior performance to n-gram models and HMMs, demonstrating that long-term historic information is important for mobility modeling and prediction. The DNN approach captures long-term regularities---e.g. if the start location of a trajectory is a dormitory, the likelihood of the trajectory ending in the same dormitory is high---whereas this information is not captured by n-gram or HMM models. 

Additionally, we observe that, due to variations in human behavior, there are errors in prediction too. For example, students frequently change the dining halls visited based on the menu at each dining hall or based on the dining location visited by their friends. Also, non-regular mobility such as visits to university health center or to an administrative office are hard to predict and the model does not capture such high variations from users' routine mobility. We also find that our model captures the recurring mobility patterns at the inter-building level with a very high accuracy of ~90\% but, due to variations introduced by human behavior such as visiting a different dining hall or carrying out an unexpected errand at an administrative building, etc results in the induction of errors. %In subsection \S \ref{sec:rec} we demonstrate a method to capture the variations across few locations that regularly show spatial variance but have almost constant temporal feature, such as visiting a different dining hall, from a set of preferred dining halls, based on the food menu but the location is visited at almost the same time everyday.
Another interesting observation is that varying the temporal scale of trajectory has an impact on the prediction accuracy. 

%\textbf{Impact of Spatial granularity} In this experiment, we vary the spatial scale of trajectory to understand the impact on model accuracy. By Spatial scale we mean location scale of trajectory as in macro scale mobility at building type or building name granularity or micro level mobility at indoor spatial granularity. We find that while macro level mobility is easier to predict due to the aggregation of all indoor mobility confined to that one building accounting to one long duration visit to a building, or space type micro level mobility is harder to predict due to large frequent transitions. 
%Temporal scale refers to the sampling rate of trajectories. We represent the user trajectory as a sequence of locations, where the location is sampled every n mins. For evaluations we use 3 sampling values of n. 15, 30, and 60 min sampling frequency resulting in trajectories labeled as T15, T30, and T60. In table \ref{tab:baseline} we find that the 60 mins sampled trajectories, T60, have the highest accuracy while T15 shows the lowest across all models. We further explore the spatio temporal efficacy in the sub-sections below. %Most of the indoor micro level prediction errors were due to unscheduled short visits that get captured in trajectories sampled every 15 mins.

\textbf{Impact of Temporal Granularity} In this experiment, we vary the temporal granularity of training and prediction. Trajectory temporal granularity refers to the sampling rate of trajectories. We represent the user trajectory as a sequence of locations, where the location is sampled every $n$ minutes. We train the model on trajectories with a temporal granularity of 15 mins, 30 mins, and 60 mins, here on referred to as T15, T30, and T60 respectively. As shown in Table \ref{tab:baseline}, we find that across all models with different sampling frequency T60 prediction is highest followed by T30 and then T15. %location predicted every 30 mins and least for trajectory with location predicted every 15 mins. 
We find that as the model temporal granularity becomes coarser, the indoor mobility accuracy increases because indoor human mobility is more frequent at fine granularity. %and becomes less frequent as the scale becomes coarser.
When we learn mobility at a coarser temporal scale of 60 mins, frequent short mobility observed at 15 min temporal scale such as a break to visit the vending machine or stop by a colleague's office for a chat gets masked. Additionally, such short micro events have high variability and cannot be predicted accurately at a fine temporal granularity resulting in reduced accuracy at a fine temporal granularity as seen in T15 trajectories ,location sampled every 15 mins.

\begin{table}[ht]
\begin{minipage}[c]{0.45\linewidth}
     \centering
    \includegraphics[scale=0.33]{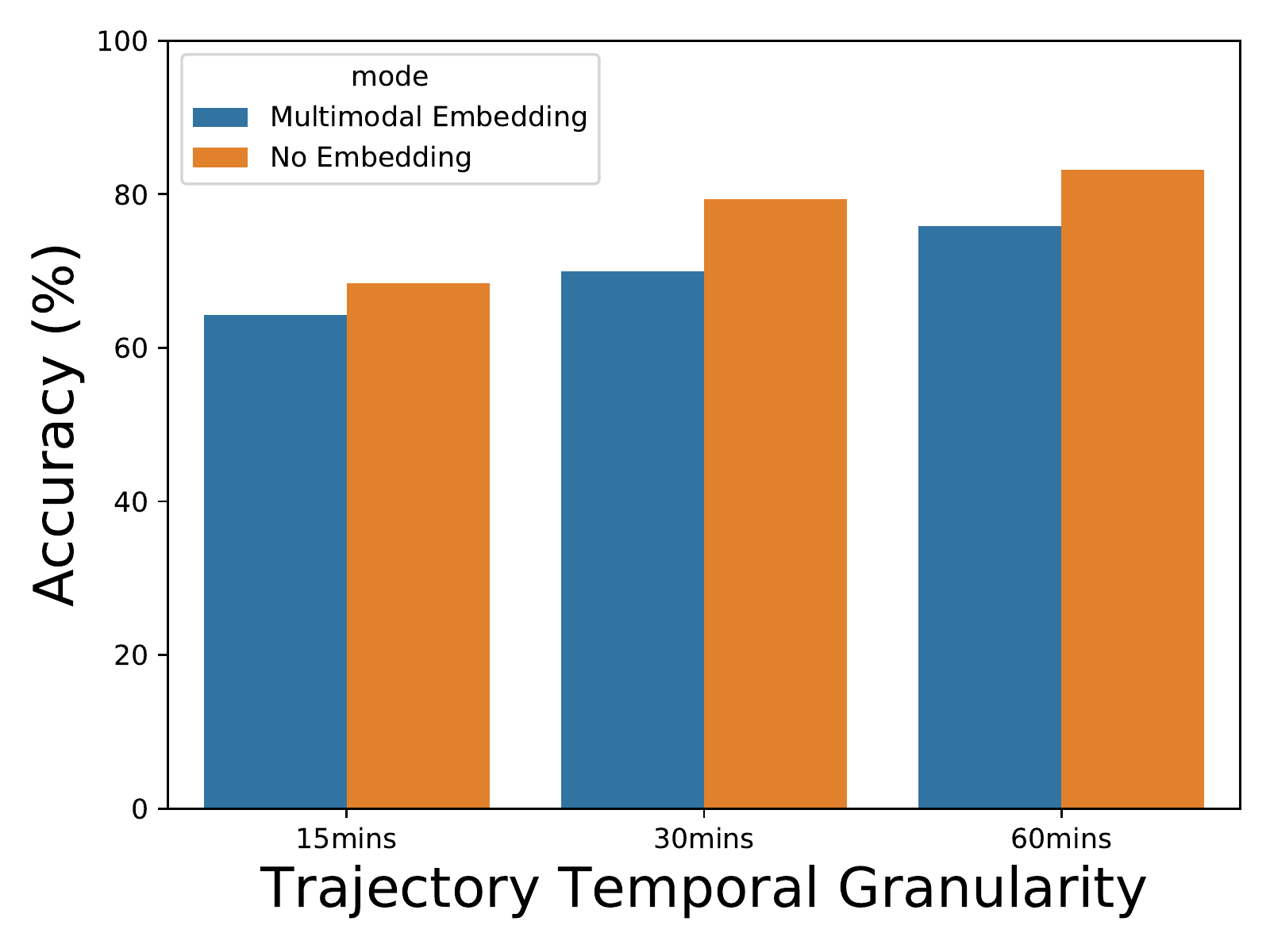}
    \captionof{figure}{Multimodal embedding effectiveness : Comparison of indoor location prediction of a transformer for multimodal and non-multimodal embedding input}
\label{fig:multimodal}
\end{minipage}
%\hfill
\begin{minipage}[c]{0.45\linewidth}
\centering
\includegraphics[scale=0.33]{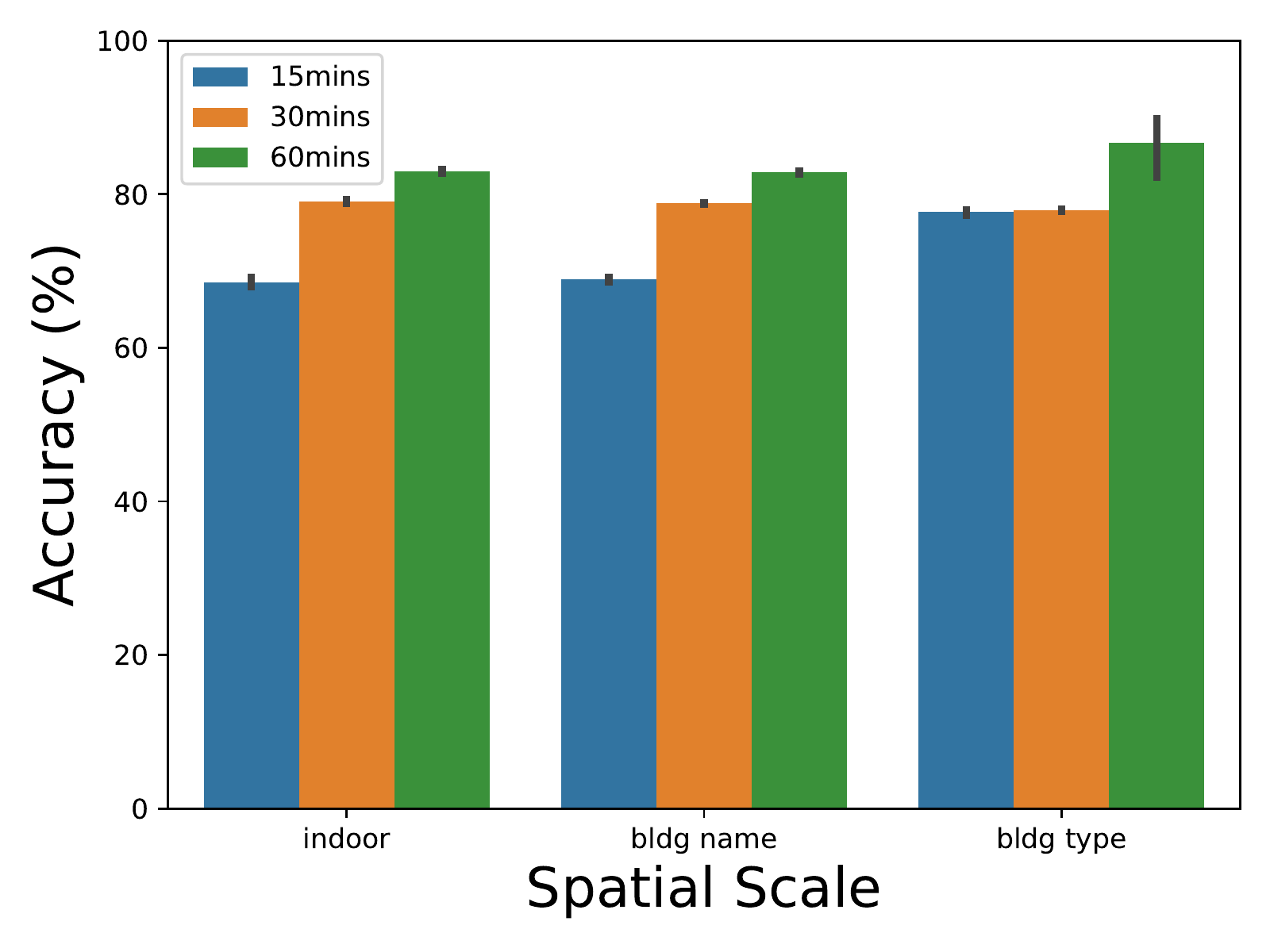}
\captionof{figure}{Space Type Prediction \label{fig:space_type}}
\end{minipage}
\end{table}

\iffalse
\begin{figure}
\centering
    \includegraphics[width=0.45\linewidth]{Figures/multimodal_embedding.pdf}
%\vspace{-0.2in}
    \caption{Multimodal embedding effectiveness : Comparison of indoor location prediction of a transformer for multimodal and non-multimodal embedding input}
\vspace{-0.05in}
\label{fig:multimodal}
\end{figure}
\fi

\subsection{Effectiveness of Multi-Modal Embedding}
In figure \ref{fig:multimodal} we see that the indoor mobility prediction accuracy of our model is higher than a single Transformer implementation that has a flat input structure of only indoor locations. To compare the two models on prediction accuracy, we predict the next top-1 location with both the models for the same test dataset on indoor location granularity. The multi-modal embedding model shows an accuracy of 83.2\% while a simple transformer with no embedding has an accuracy of 75.81\% for T60 trajectories. The multi-modal embedding approach outperforms the non-embedding approach even for T15 and T30 trajectories demonstrating that modeling mobility from a hierarchical perspective where the model learns the correlations across multiple spatial scale mobility using multi-modal embedding results in higher prediction accuracy. The intuition behind higher accuracy is that the multi-modal approach significantly reduces the prediction space by learning the correlations between macro and micro scale mobility, conditioning the prediction on the estimated macro and micro scale mobility distribution, thereby using the topological constraint of the multiple spatial scales. This behaviour is also reflected in Table \ref{tab:hierarchy_flat}, which compares the accuracy of hierarchical and non-hierarchical model across multiple spatial scales. Additionally, the model also captures the correlation and periodicity in mobility across varying spatial scales.

\begin{table}
    \begin{center}
     \begin{tabular}{ccc} \toprule
     Spatial Scale & \sysname & Non-Hierarchical \\ \midrule
     Building Type & 89.58\% &  89.23\%\\
     Building Name & 87.39\% & 81.12\%\\
     Indoor Location & 83.2\% & 75.81\% \\ \bottomrule
     \end{tabular}
     \caption{Comparison of accuracy of hierarchical and non-hierarchical model across multiple spatial scale.}
     \label{tab:hierarchy_flat}
 \end{center}
\vspace{-0.3in}
\end{table}

\iffalse
\begin{figure}[h]
\centering
\includegraphics[scale=0.45]{Figures/space_vs_sample.pdf}
\caption{Space Type Prediction \label{fig:space_type}}
\end{figure}
\fi

\subsection{Importance of Space Type Prediction}
While analyzing the indoor location predictions made by the model, we find that most of the errors are in predicting food court location and space inside food courts, indoor library locations of use, indoor location inside the recreation center, etc. These locations have a high variance when predicting the indoor location. However, we find that the model displays a high accuracy in predicting the context, and location types and low accuracy on building name, in the case of multiple food courts, or indoor location of use. Figure \ref{fig:space_type} analyzes the model accuracy by space type. We see that the model has high prediction accuracy for building type followed by building name and lowest for indoor location across all 3 sampling frequencies T15, T30, and T60. This is mainly because, while routine activities such as visiting the classrooms, office space, research labs have fixed building type, building name, and indoor location while visits to high variance locations such as library or dining hall has a fixed building type but variance in indoor locations (since the person might not sit at the same location always) and building name (since the person might not visit the same building under the building type, such as dining hall). Additionally, we find that most errors are found in indoor location prediction, fine spatial granularity for high sampling rate of 1 sample every 15 mins in T15 because this trajectory captures the most unscheduled high variance micro mobility at a fine spatial scale.

\begin{figure}
\centering
\includegraphics[scale=0.33]{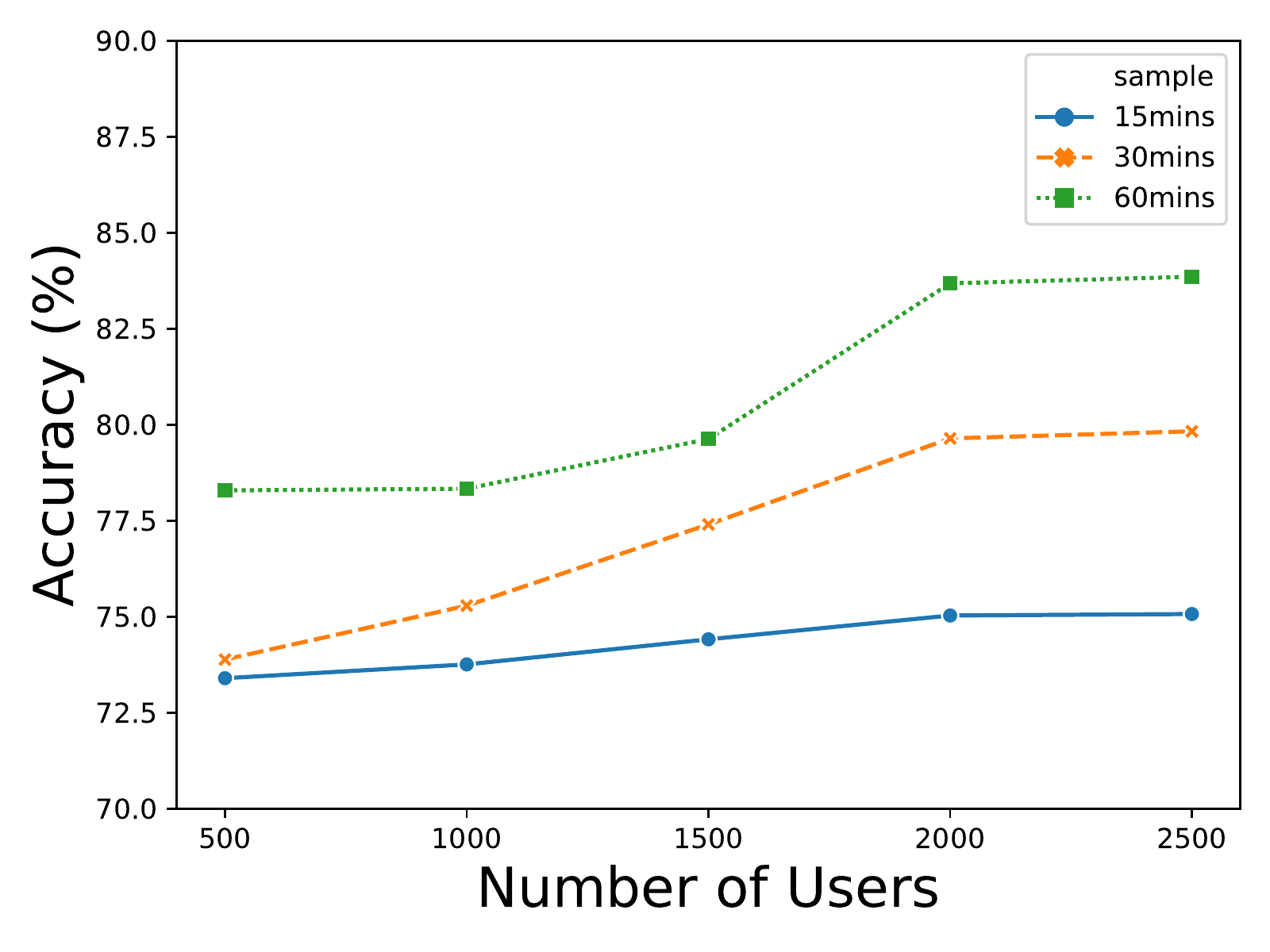}
\caption{Impact of number of users on model accuracy \label{fig:user_count}}
\end{figure}

%\textbf
\subsection{Impact of the number of trajectories} We vary the training dataset by using a subset of trajectories, with subset sizes of 500,1000,1500,2000, and 2500 user trajectories. We train the models on the subset of user trajectories for the first 7 weeks of the semester and predict the user trajectories for the next 2 weeks. We find that the transformer based model displays higher accuracy for larger training set size indicating that the model has better generalizability and higher performance for more and new data. The model accuracy for T15 increases the most from 73.4\% to 75.06\%, whereas model accuracy for T60 increases from 78.28\% to 83.2\% . Across all trajectories, with different temporal binning we see that the model accuracy increases as we increase the number of user trajectories in the training dataset.

\iffalse
\begin{figure}
\includegraphics[scale=0.45]{Figures/tier1_topK_predictions_accuracy.pdf}
\caption{Tier-1 top 1 and top 5 predictions of locations made by the model. \label{fig:topk}}
\end{figure}
\fi

%\textcolor{red}{Amee working on this subsection}
%\textbf{Impact of Prediction Horizon length} We vary the horizon of prediction from next location, half day trajectories to entire day trajectories and find that as we increase the prediction horizon the accuracy decreases. However, the accuracy drop slightly from xx to yy for faculty and staff users who have an almost fixed schedule while model accuracy drops from aa to bb for students over the course of the entire day. We see that the drop is significant after p steps when they usually visit building type that has   based on the time of the day accuracy decreases. Put the plot and explain the results.
\section{Case Studies}
\label{sec:casestudy}

\begin{figure}[ht]
\begin{center}
    \begin{tabular}{ccc}
    \includegraphics[height=0.65in]{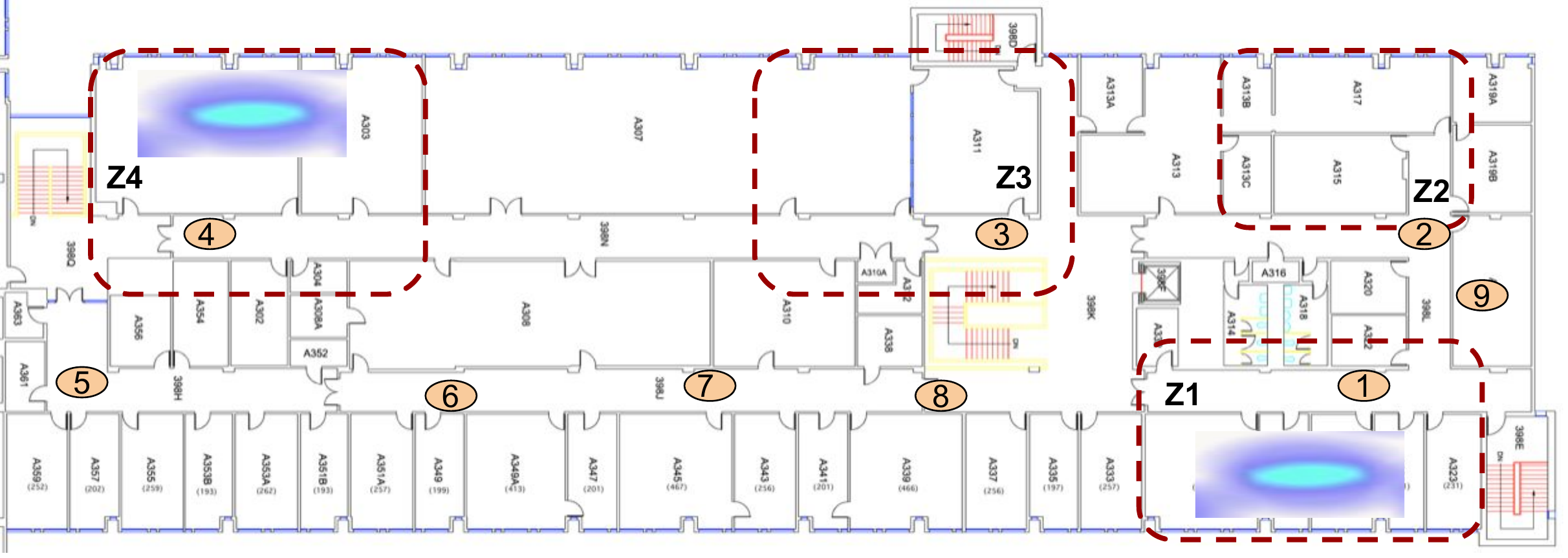} &
    \includegraphics[height=0.65in]{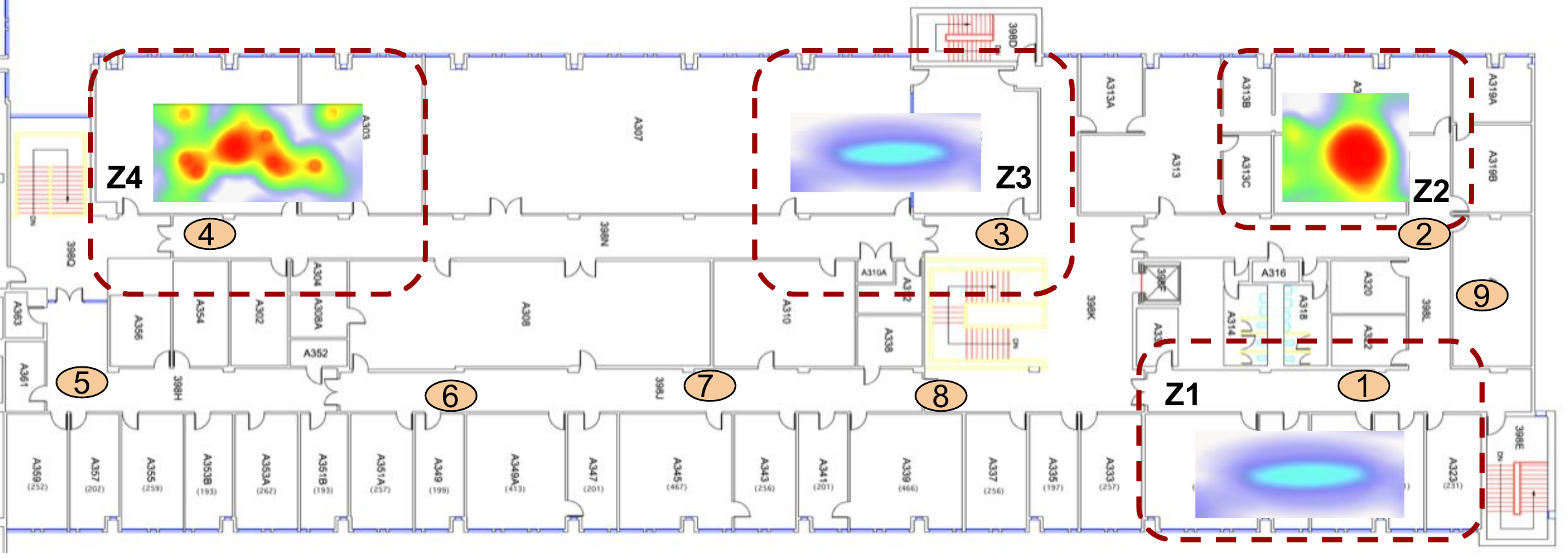} &
    \includegraphics[height=0.65in]{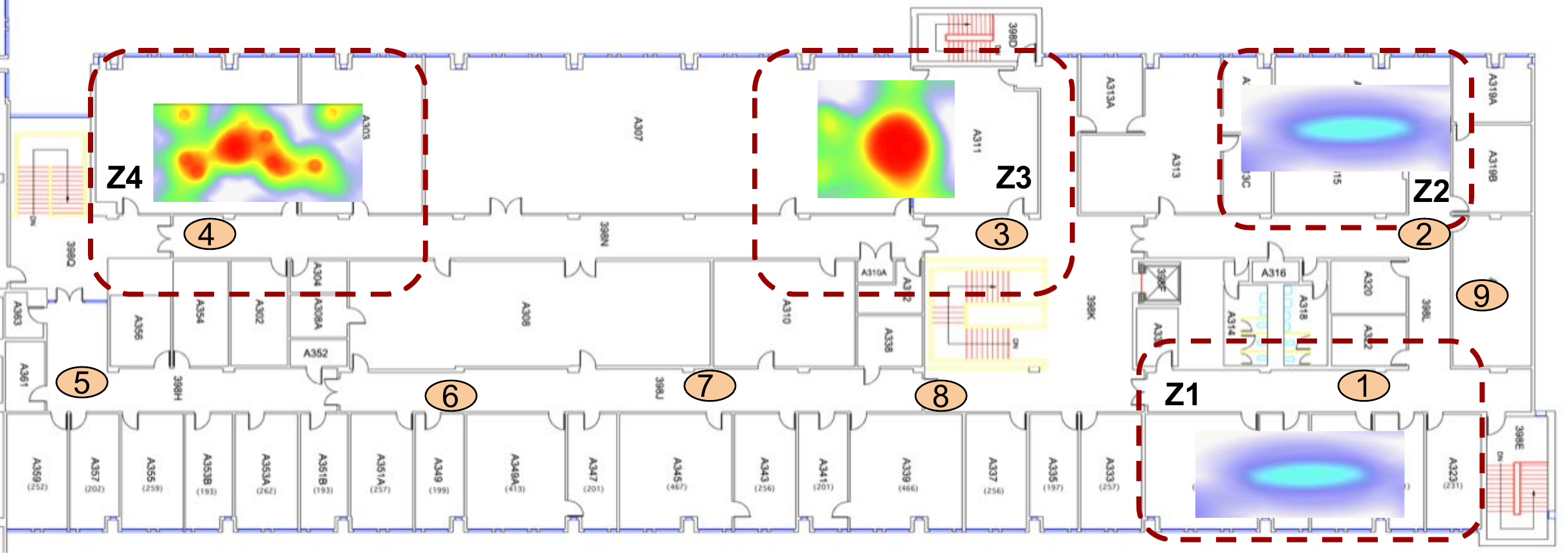}\\
(a) 8:30am & (b) 12:00pm(Noon) & (c) 3:30pm
\end{tabular}
\end{center}
%\vspace{-0.2in}
    \caption{Heatmap of predicted indoor occupancy of educational bldg with classrooms, research labs, faculty office, kitchenette}% r\textsuperscript{2} score : 0.989 and  0.984  respectively.}
%\vspace{-0.05in}
\label{fig:occ_pred}
\end{figure}

In this section we discuss three case studies of our proposed system \sysname.

\subsection{Casestudy 1: Indoor COVID19/ILI Hotspot Prediction}

%\begin{itemize}
%    \item Space utilization changes changes throughout the day
%    \item we predict occupancy across zones from the global model
%    \item predicted and observed are very close - accuracy of xxx\%
%\end{itemize}

With the current COVID-19 pandemic, building occupancy scheduling and resource allocation for de-densification is a key component in designing re-opening policies. Here, we present a case study of using \sysname to predict indoor mobility of a building across the coarse of an entire day(s) to identify indoor spaces with high space utilization that can become a hot pocket zone and needs dedensification so that the number of users inside the building is always below the 50\% or 25\% usage constraint for space usage..% This casestudy and \sysname model usage can also serve as a building block for generating schedules and policies that adhere to de-densification of indoor occupancy %. This casestudy can be used for designing policies for restricting the number of users across all rooms, zones, floors of the building for restricting the number of users across all rooms, zones, floors of the building so that the number of users inside the building is always below the 50\% or 25\% usage constraint for space usage.

Here, we use the trained model to predict the user trajectories of all users on campus for the entire day. We then aggregate all these predicted trajectories across the temporal attribute to compute the occupancy at each indoor location at various times of the day. In our case since we are using WiFi AP syslogs, the indoor spatial granularity is zone level where each AP captures occupancy per zone that might encapsulate a single room or across few rooms, based on the range of AP deployed. Fig \ref{fig:occ_pred} shows the floor map of an educational building with 9 deployed APs. The floor has a combination of faculty office, break room, research labs, and classrooms. We focus on APs 1-4 which have a range across zones Z1-Z4 respectively as indicated in figure \ref{fig:occ_pred}. Zone Z1 encompasses few faculty offices and a research lab, Z2 spans across a kitchenette and a research lab, Z3 across a conference room and a student work space, and Z4 across a classroom and a research lab. Fig \ref{fig:occ_pred} (a)-(d) shows the computed indoor user occupancy based on the model predictions, at 3 different times of the day. We find that at 8:30am fig \ref{fig:occ_pred}(a) the space occupancy is very low with the start of the day across all zones, with some occupancy in Z1 and Z4. Fig \ref{fig:occ_pred}(b) shows predicted space occupancy at noon and we observe high human density across Z4(classroom zone) with an in-person class (in 2019), Z2(kitchenette area) with the break room where students gather to eat lunch, and the rest zones show low to moderate occupancy. Fig \ref{fig:occ_pred}(c) shows predicted space usage at 3pm and we see that zones Z3, and Z4 have high occupancy due to predicted recurring seminar, and classroom usage respectively while Z2 kitchennet and Z1 lab space have relatively low occupancy. Fig \ref{fig:occ_pred}(d) shows predicted space usage at 5pm and we see some occupancy in zones Z1 and Z2 that comprise of research labs with students still working in late evenings. The computed occupancy across the 3 times of the day shows an accuracy of 96\% as compared with observed ground truth indoor occupancy computed from WiFi logs.
%Fig \ref{fig:occ_obs}(a)-(d) show the actual occupancy as derived from WiFi logs and we see a very high accuracy of 96\% between the predicted and ground truth observed indoor occupancy.

In the heatmaps \ref{fig:occ_pred}(a)-(c) the red areas indicate high human density or hotpocket zones on the floor map. We can generate such heatmaps for all indoor spaces across the times of the day to identify hotpockets and design policies or space usage schedules to de-densify them to lower the risk of disease spread and safe opening of indoor spaces.

Additionally, indoor location occupancy computed by aggregating indoor mobility can also be used to generate customized Heating, Ventilation, and Air Conditioning (HVAC) schedules per building. Such customized HVAC schedules can help reduce the energy consumption while increasing the user comfort by scheduling HVAC to turn-on with predicted indoor occupancy while turning it off with low to no indoor occupancy.

\begin{figure}[H]
\begin{center}
    \begin{tabular}{cc}
    \includegraphics[width=0.33\linewidth]{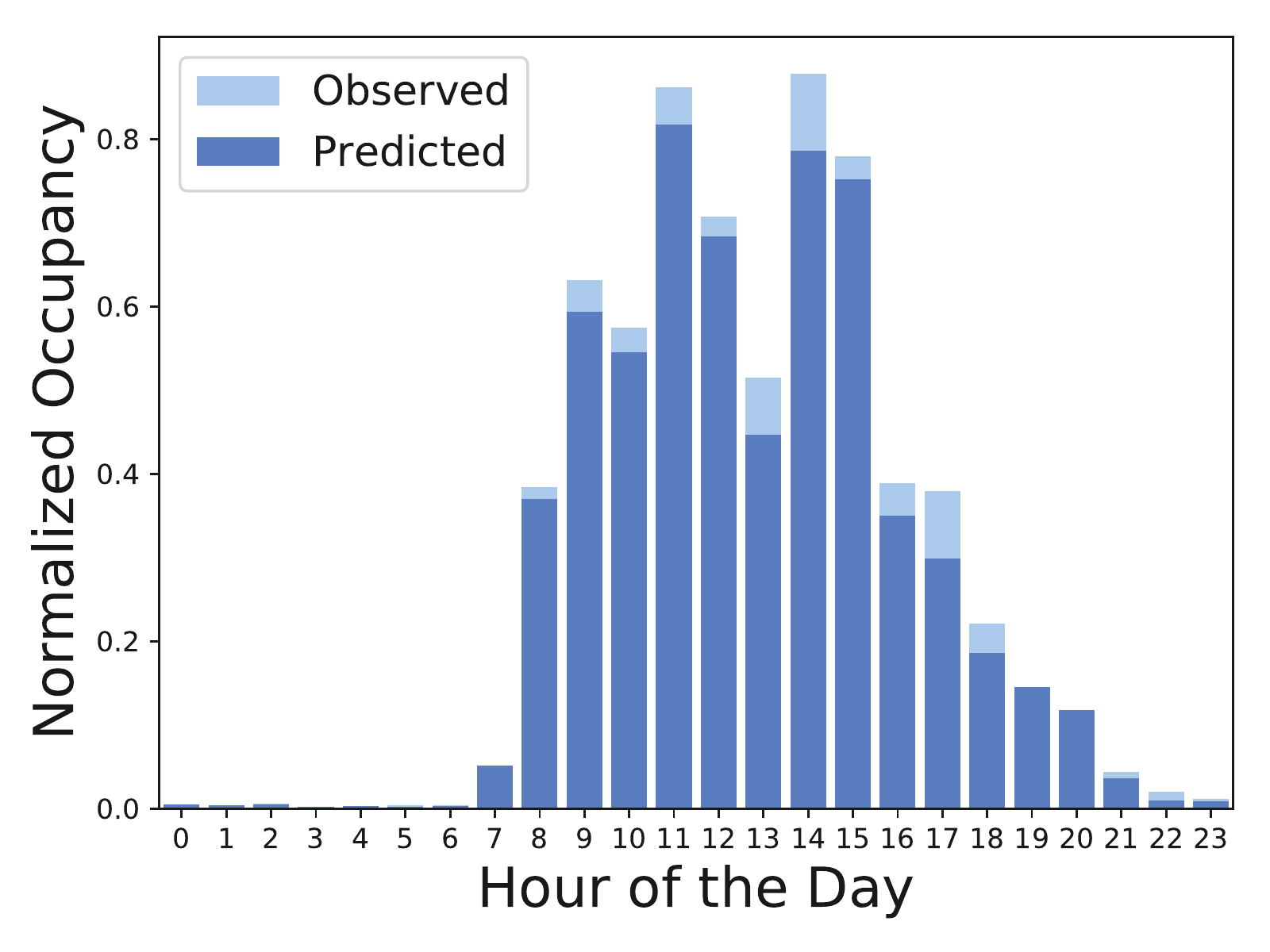} &
    \includegraphics[width=0.33\linewidth]{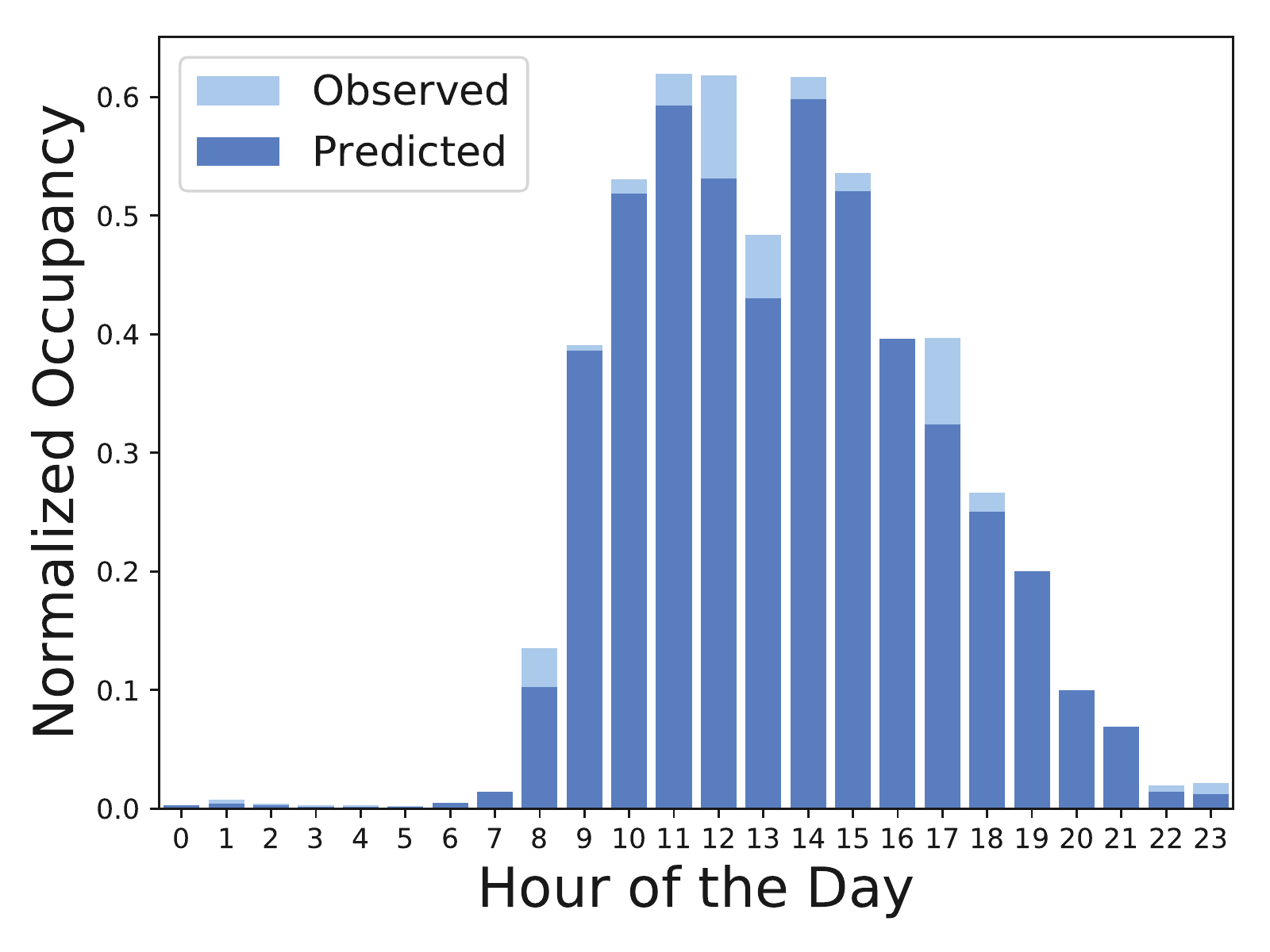}\\
(a) Location1(r\textsuperscript{2} 0.989) & (b) Location2(r\textsuperscript{2} 0.984)

\end{tabular}
\end{center}
%\vspace{-0.2in}
    \caption{Normalized hourly occupancy across 2 locations computed from actual and simulated trajectories.}% r\textsuperscript{2} score : 0.989 and  0.984  respectively.}
\vspace{-0.05in}
\label{fig:occ_count}
\end{figure}

\subsection{Casestudy 2: Human Mobility Simulation}
Mobility datasets are fundamental to  evaluation of a system or applications such as simulation of disease spread. However, such datasets are hard to obtain due to privacy concerns. Majority of the mobility trajectory generators use deterministic models that have predefined mobility distributions or assume human mobility follows levy walk, random distribution or stochastic process, failing to capture realistic mobility. This leads to a gap in analyzing and fine tuning systems. Hence, we propose scenario simulation by generating synthetic mobility trace using our pre-trained hierarchical model. To demonstrate the efficacy of our model for synthetic trace generation we generate the trajectory of users and their devices using a pre-trained model on the campus mobility dataset. We compare the hourly occupancy at 2 different locations computed from the synthetic trace and observed trajectories. The simulation is performed at 20\% of the total population, and the observed transition and occupancy are scaled down accordingly. Figure ~\ref{fig:occ_count}(a) and (b) compares the hourly occupancy of 2 different locations, loc1 and loc2, and the model demonstrates a high accuracy with the coefficient of determination, r\textsuperscript{2} value as 0.989 and 0.984 for the loc1 and loc2 respectively.
%We compare the results of (i) volume of transitions across campus over the course of the day (ii) hourly occupancy at 2 different locations (Dining and Education building) computed from the synthetic trace and observed trajectories. The simulation is performed at 20\% of the total population, and the observed transition and occupancy is scaled down accordingly. Figure ~\ref{fig:} compares the number of people transitioning through the campus over the course of an entire day and we see a similarity of x\% between the synthetic transition time distribution and observed distribution. Figure ~\ref{fig:} compares the hourly occupancy of 2 different locations on campus (Dining and an Educational building), it again shows a high accuracy of X\%.

For applications such as user profiling and behaviour analysis, which need to capture variations in human behavior we can introduce realism in capturing the variations in human behavior by changing the inference mechanism in the decoder from selecting the next location that gives least negative log-likelihood to sampling the next location from the top-5 possible next predictions. To validate if our synthetic traces are close to the real dataset, we do a domain search of the generated trace to actual observed trace and find trajectory similarity score of 82\% on weekdays and 63\% on weekends for indoor mobility.

\subsection{Casestudy 3: Single User Personal Assistant} 
Last few years have seen an introduction of personal assistants that share the goal of presenting the user the right information at the right time. However, knowing when to present the information without any query from the user is an important criterion and a critical limitation in many of today's models. Since, the information presented is mainly associated with current location, time of day, space type, and user type. We propose that a user mobility model derived by using \sysname serves as a foundation for a highly accurate personal assistant that can be used for informing users with a variety of tasks/events/updates. We use a globally trained model and fine tune it for each user by locally training it with the historic trajectories of each user to create a personalized model and use it to make macro and micro scale predictions. We find that our model shows high indoor mobility prediction accuracy in the top-1(accuracy of most likely location) prediction score is ~89\% for user type faculty/staff and ~85\% for user type student for weekdays. Such a model can be augmented with the user calendar or campus event calendar to notify the user with upcoming events of interest or prior scheduled classes or meetings.

%The primary device trajectory is a very close to reality mobility trajectory of the user since mobile phones are carried everywhere by the users and use the personalized models to send notifications to the user for the upcoming events learned from historic mobility.  We further add notification messages that appear 5 mins before the transitions in mobility are predicted along with the destination location. If traffic information is provided as an input to our system we can also suggest alternative transition paths. There are multiple features such as event calendars, LSBN notifications or context of each trip that can be added to enhance our personal assistant.

%Compare independent and conditional device multi device mobility modeling.

%\subsection{Embedding Visualizations}
%Use t-SNE for dimensionality reduction and plot the embedding at:
%\begin{itemize}
%    \item Inter building
%    \item Intra building 
%    \item Device mobility 
%    \item Describe the results and take aways
%\end{itemize}

\section{Related Work}
\label{sec:related_work}

%\textbf{Deep-Learning based Models :} 
%One body of work has employed Markov Models or Hidden Markov Models (HMMs) to capture the sequential nature of human mobility \cite{chen2014nlpmm, gambs2012next, Ganti:2013:IHM:2493432.2493466, Lee:2006:MST:1132905.1132915}. In such models, each node represents a unique location and edges denote transition probabilities from one location to another. Such approaches have been shown to provide better accuracy than early methods that modeled mobility as a Random walk (e.g, Continuous Time Random Walk). However, capturing long-term dependence in the data or recurring patterns is challenging when using Markov Models; doing so requires the use of higher-order Markov models, which quickly grow in complexity and computational overheads.

There has been significant work on using Markov Models or Hidden Markov Models (HMMs) to capture the sequential nature of human mobility. %Such models outperform the early methods of Random walk. 
However, capturing long-term dependence in the data or recurring patterns is challenging when using Markov Models; doing so requires the use of higher-order Markov models, which quickly grow in complexity and computational overheads.

Most of the mobility modeling work focusses on outdoor mobility modeling at urban-scales \cite{isaacman2012human, lin2017deep, Lee:2006:MST:1132905.1132915, kim2006extracting} %\cite{zhao2016urban, xia2018exploring,isaacman2012human,d2017if}
, next location prediction \cite{Do:2012:CCM:2370216.2370242, liu2016predicting, Lin:2012:PIM:2370216.2370274, gambs2012next, gidofalvi2012and, mathew2012predicting}, and point of interest areas \cite{yuan2012discovering} using a variety of data sources such as  cellular, WiFi, social media check-ins, and vehicular data \cite{Ganti:2013:IHM:2493432.2493466, hang2018exploring, veloso2011urban, hasan2013understanding, jurdak2015understanding}. All these outdoor models cater to a discrete mobility models where mobility is infrequent compared to fine grain indoor mobility hence %We argue that models designed for outdoor mobility are 
these outdoor models cannot be directly applied to indoor environments.% since indoor mobility models need to address a different set of considerations. Our work leverages the state-of-the-art techniques from NLP while using passive sensing to provide a reliable, cheap, and easy to deploy data-driven technique to model indoor human mobility. %and cIn contrast, to extensive work on outdoor mobility, there is relatively little work on indoor mobility models.

More recent work in this area has focused on urban mobility modeling using cellular, transportation or social media data using data driven methods, specifically deep learning. Recurrent Neural Networks (RNNs) have emerged as a popular approach for urban mobility modeling \cite{lin2017deep, Feng:2018:DPH:3178876.3186058, song2016deeptransport, zhang2018deeptravel, jiang2018deep} taking inspiration from Natural Language Processing (NLP) to learn long term dependencies. ST-RNN model \cite{al2016stf} models spatial and temporal contexts of mobility but is too complicated with the need to tune a lot of parameters and cannot be easily deployed for indoor mobility which is very frequent. In DeepMove \cite{Feng:2018:DPH:3178876.3186058} the authors propose using RNN to model sparse trajectories. It does not cater to either indoor mobility or capturing the mutli-scale hierarchical mobility correlations. 

Other efforts in indoor mobility modeling comprise of \cite{jayarajah2018predicting} but this approach is for modeling mobility based on groups and social friendship ties. Additionally they use aaa to acquire the dataset and it requires human effort and cost in acquisition. \sysname uses passively sensed WiFi syslogs and doesn't need any new infrastructure or human feedback for data collection. There has been work on using WiFi probes for sensing where the probe requests from mobile devices that steadily scan the APs close by for access are used for monitoring the crowds or activity flow for monitoring users \cite{xi2014electronic, weppner2016monitoring}. These prior works do not focus on individual mobility and instead look at crowd and activity behavior of aggregated users. In our work we focus on individual user mobility trajectories using WiFi syslogs and not WiFi probes. Additionally, our model supports all applications that need individual as well as aggregated human mobility unlike only aggregated behavior as analyzed by prior work.

\section{Conclusions}
\label{sec:conclusions}

Modeling indoor mobility and using the correct spatial granularity of mobility can substantially benefit a large range of applications. In this paper, we proposed \sysnames, a data-driven approach to model indoor human mobility using passively sensed WiFi logs.  In \sysname we jointly model mobility context, space type, outdoor location, and indoor location using a transformer to learn the correlations of mobility at various spatial granularities. We extensively evaluated our approach using available ground truth WiFi data from 2500 users at a large university campus and found that our model outperforms the current state-of-the-art baselines significantly.  Further, we also demonstrated the need and use of modeling mobility at multiple spatial scales. Additionally, we demonstrated that our proposed approach can be applied to many other real-world applications such as Personal assistant design, trajectory simulation, and indoor human density or hot pocket prediction to help with resource allocation, scheduling, as well as COVID19 de-densification policy compliance among many others.

\bibliographystyle{ACM-Reference-Format}
\bibliography{paper,paper_ks_additions}

%%
%% If your work has an appendix, this is the place to put it.
%\appendix

\end{document}